\documentclass[12pt]{article}

\usepackage{graphicx}
\usepackage{float}
\usepackage{cite}
\usepackage{amsfonts}
\usepackage{amssymb}
\usepackage{amsmath}

\def\gtwid{\mathrel{\raise.3ex\hbox{$>$\kern-.75em\lower1ex\hbox{$\sim$}}}}
\def\ltwid{\mathrel{\raise.3ex\hbox{$<$\kern-.75em\lower1ex\hbox{$\sim$}}}}
\def\square{\kern1pt\vbox{\hrule height 1.2pt\hbox{\vrule width 1.2pt\hskip 3pt
   \vbox{\vskip 6pt}\hskip 3pt\vrule width 0.6pt}\hrule height 0.6pt}\kern1pt}

\begin{document}

\begin{titlepage}

\begin{flushright}
UFIFT-QG-19-05 , CCTP-2019-8
\end{flushright}

\vskip 2cm

\begin{center}
{\bf The Inflaton Effective Potential for General $\epsilon$}
\end{center}

\vskip 1cm

\begin{center}
A. Kyriazis$^{1*}$, S. P. Miao$^{2\star}$, N. C. Tsamis$^{1\dagger}$ and R. P. Woodard$^{3\ddagger}$
\end{center}

\vskip .5cm
\begin{center}
\it{$^{1}$ Institute of Theoretical Physics \& Computational Physics, \\
Department of Physics, University of Crete, \\
GR-710 03 Heraklion, HELLAS}
\end{center}

\begin{center}
\it{$^{2}$ Department of Physics, National Cheng Kung University, \\
No. 1 University Road, Tainan City 70101, TAIWAN}
\end{center}

\begin{center}
\it{$^{3}$ Department of Physics, University of Florida,\\
Gainesville, FL 32611, UNITED STATES}
\end{center}

\vspace{1cm}

\begin{center}
ABSTRACT
\end{center}
We develop an analytic approximation for the coincidence limit of a 
massive scalar propagator in an arbitrary spatially flat,
homogeneous and isotropic geometry. We employ this to compute the one
loop corrections to the inflaton effective potential from a quadratic
coupling to a minimally coupled scalar. We also extend the Friedmann
equations to cover potentials that depend locally on the Hubble 
parameter and the first slow roll parameter.

\begin{flushleft}
PACS numbers: 04.50.Kd, 95.35.+d, 98.62.-g
\end{flushleft}

\vskip .5cm

\begin{flushleft}
$^{*}$ e-mail: ph4845@edu.physics.uoc.gr \\
$^{\star}$ e-mail: spmiao5@mail.ncku.edu.tw \\
$^{\dagger}$ e-mail: tsamis@physics.uoc.gr \\
$^{\ddagger}$ e-mail: woodard@phys.ufl.edu
\end{flushleft}

\end{titlepage}

\section{Introduction}

Certain models of scalar-driven inflation,
\begin{equation}
\mathcal{L} = \frac{R \sqrt{-g}}{16 \pi G} - \frac12 \partial_{\mu} \varphi
\partial_{\nu} \varphi g^{\mu\nu} \sqrt{-g} - V(\varphi) \sqrt{-g} \; .
\label{scalarinfl}
\end{equation}
are still consistent with the data from cosmological perturbations 
\cite{Aghanim:2018eyx}. However, these models must be strongly fine-tuned in 
order to make inflation start, to make the potential nearly flat, and to remain 
predictive by avoiding the formation of a multiverse \cite{Ijjas:2013vea}. The 
resulting controversy \cite{Guth:2013sya,Linde:2014nna,rebuttal} within the 
community of inflationary cosmologists has been described as a schism 
\cite{Ijjas:2014nta}.

We have become worried about another sort of fine-tuning problem associated
with coupling the inflaton to ordinary matter in order to make re-heating 
efficient. As always with such a coupling, the 0-point motion of quantum matter 
engenders Coleman-Weinberg \cite{Coleman:1973jx} corrections to the inflaton 
potential. These corrections are dangerous to inflation because they are not
Planck-suppressed \cite{Green:2007gs}. Nor can they be completely subtracted 
off by allowed counterterms because they depend in a complex way on the Hubble 
parameter for de Sitter \cite{Miao:2015oba}, which is the only background on 
which they have been computed. If one simply assumes that the constant Hubble 
parameter of de Sitter becomes the instantaneous Hubble parameter in the 
evolving geometry of inflation then there are two allowed subtraction schemes:
\begin{enumerate}
\item{Remove a function of just the inflaton \cite{Liao:2018sci}; or}
\item{Remove a function of the inflaton and the Ricci scalar \cite{Miao:2019bnq}.}
\end{enumerate}
Neither of these schemes leads to acceptable results, and no more general 
metric dependence is permitted by locality, invariance and stability  
\cite{Woodard:2006nt}.

The weak point in these studies is the assumption that the constant Hubble 
parameter of de Sitter computations goes over to the instantaneous Hubble 
parameter of inflation. Our purpose in this paper is to understand how 
$V_{\rm eff}(\varphi)$ depends on the geometry of inflation,
\begin{equation}
ds^2 = -dt^2 + a^2(t) d\vec{x} \!\cdot\! d\vec{x} \qquad \Longrightarrow \qquad
H \equiv \frac{\dot{a}}{a} \quad , \quad \epsilon \equiv -\frac{\dot{H}}{H^2} \; .
\label{FLRW}
\end{equation}
In this geometry the inflaton $\varphi_0(t)$ depends only on time and the 
nontrivial Einstein equations are,
\begin{eqnarray}
3 H^2 & = & 8\pi G \Bigl[ \frac12 \dot{\varphi}_0^2 + V(\varphi_0) \Bigr] \; , 
\label{Infl1} \\
- (3 \!-\! 2 \epsilon) H^2 & = & 8 \pi G \Bigl[ \frac12 \dot{\varphi}_0^2 - 
V(\varphi_0) \Bigr] \; . \label{Infl2}
\end{eqnarray}
The inflaton itself evolves according to the equation,
\begin{equation}
\ddot{\varphi}_0 + 3 H \dot{\varphi}_0 + V'(\varphi_0) = 0 \; . \label{Infl3}
\end{equation}

For definiteness, we couple the inflaton $\varphi$ to a massless, minimally 
coupled scalar $\phi$, with conformal and quartic counterterms,
\begin{equation}
\Delta \mathcal{L} = -\frac12 \partial_{\mu} \phi \partial_{\nu} \phi g^{\mu\nu} 
\sqrt{-g} - \frac{h^2}{4} \phi^2 \varphi^2 \sqrt{-g} -\frac{\delta \xi}{2} \varphi^2 
R \sqrt{-g} - \frac{\delta \lambda}{4!} \varphi^4 \sqrt{-g} \; . \label{scalarcoupling}
\end{equation}
Then the derivative of the one loop correction to the inflaton potential 
$V_{\rm eff}(\varphi)$ can be expressed in terms of the coincidence limit of the
$\phi$ propagator in the inflationary background,
\begin{equation}
\frac{\partial V_{\rm eff}}{\partial \varphi} = \Biggl[ \delta \xi R + \frac16 
\delta \lambda \varphi^2 + \frac12 h^2 i\Delta(x;x) \Biggr] \varphi \; . 
\label{Veffeqn}
\end{equation}
This coincidence limit can be expressed as the inverse Fourier transform (regulated 
in $D$ spacetime dimensions) of the $\phi$ plane wave mode functions $u(t,k,M)$,
\begin{equation}
i\Delta(x;x) = \int \!\! \frac{d^{D-1}k}{(2\pi)^{D-1}} \, \Bigl\vert u(t,k,M)
\Bigr\vert^2 \; . \label{Ftrans}
\end{equation}
The mode functions obey the equations,
\begin{equation}
\ddot{u} + (D \!-\! 1)  H \dot{u} + \Bigl(M^2 + \frac{k^2}{a^2}\Bigr) u = 0 \quad , 
\quad u \dot{u}^* - \dot{u} u^* = \frac{i}{a^{D-1}} \; , \label{ueqns}
\end{equation}
where the $\phi$ mass is $M^2 = \frac12 h^2 \varphi_0^2$. Hence we seek to study 
how the amplitude of a massive scalar mode function depends upon the geometry of 
inflation.

In section 2 we numerically evaluate $\vert u(t,k,M)\vert^2$ for a simple model of
inflation to identify three distinct phases of evolution. We also demonstrate the 
validity of analytic approximations for these phases. In section 3 we apply the 
analytic approximations to compute the coincident propagator (\ref{Ftrans}) and
fully renormalize the effective potential. The resulting potential mostly depends
locally on the instantaneous Hubble and first slow roll parameters but also has a
small nonlocal part. Section 4 extends the usual Friedmann equations to cover 
Lagrangians which depend locally on $H$ and $\epsilon$. Our conclusions comprise
section 5.

\section{Approximating the Amplitude}

This section is the heart of the paper. In it we first change from co-moving time to
the number of e-foldings and re-scale the various parameters to make them dimensionless,
then evolution equations are given for the inflationary geometry and for the logarithm
of the norm-squared mode function. Next we survey the three phases of evolution, and
graphically demonstrate the validity of approximate functional forms. The section closes
with an analytic derivation of the approximations. 

\subsection{Dimensionless Evolution Equation}

It is best to measure time using the number of e-foldings $n \equiv \ln[a(t)/a(t_i)]$ 
since the start of inflation at $t = t_i$. The variable $n$ is preferable to $t$ both 
because $n$ is dimensionless and because it is less sensitive to dramatic changes which 
take place in the time scale of events as inflation progresses \cite{Finelli:2008zg}. 
Derivatives obey,
\begin{equation}
\frac{d}{dt} = H \frac{d}{dn} \qquad , \qquad \frac{d^2}{d t^2} = H^2 \Bigl[ 
\frac{d^2}{d n^2} - \epsilon \frac{d}{d n} \Bigr] \; . \label{timeton}
\end{equation}
Just as dots denote differentiation with respect to $t$ we use primes to stand for
differentiation with respect to $n$.\footnote{This only applies to functions whose
natural argument is $n$. For the potential $V(\varphi)$ we continue to employ the
prime to denote differentiation with respect to $\varphi$.} It is convenient to factor 
the dimensions out of the inflaton, the Hubble parameter and the inflaton potential,
\begin{equation}
\psi(n) \equiv \sqrt{8 \pi G} \, \varphi_0(t) \quad , \quad \chi(n) \equiv \sqrt{8\pi G} 
\, H(t) \quad , \quad U(\psi) \equiv (8\pi G)^2 V(\varphi_0) \; . \label{dimless1}
\end{equation}
Of course the first slow roll parameter $\epsilon(n) = -\chi'/\chi$ is already 
dimensionless. Using these variables we can re-express the scalar evolution equation 
(\ref{Infl3}) as,
\begin{equation}
\psi'' + (3 \!-\! \epsilon) \psi' + \frac{U'(\psi)}{\chi^2} = 0 \; . \label{Infl4}
\end{equation}
And the geometrical quantities follow from (\ref{Infl1}-\ref{Infl2}),
\begin{equation}
\chi^2 = \frac{U}{3 \!-\! \frac12 {\psi'}^2} \qquad , \qquad \epsilon = \frac12 
{\psi'}^2 \; . \label{Infl5}
\end{equation}

\begin{figure}[H]
\includegraphics[width=4.75cm,height=4.75cm]{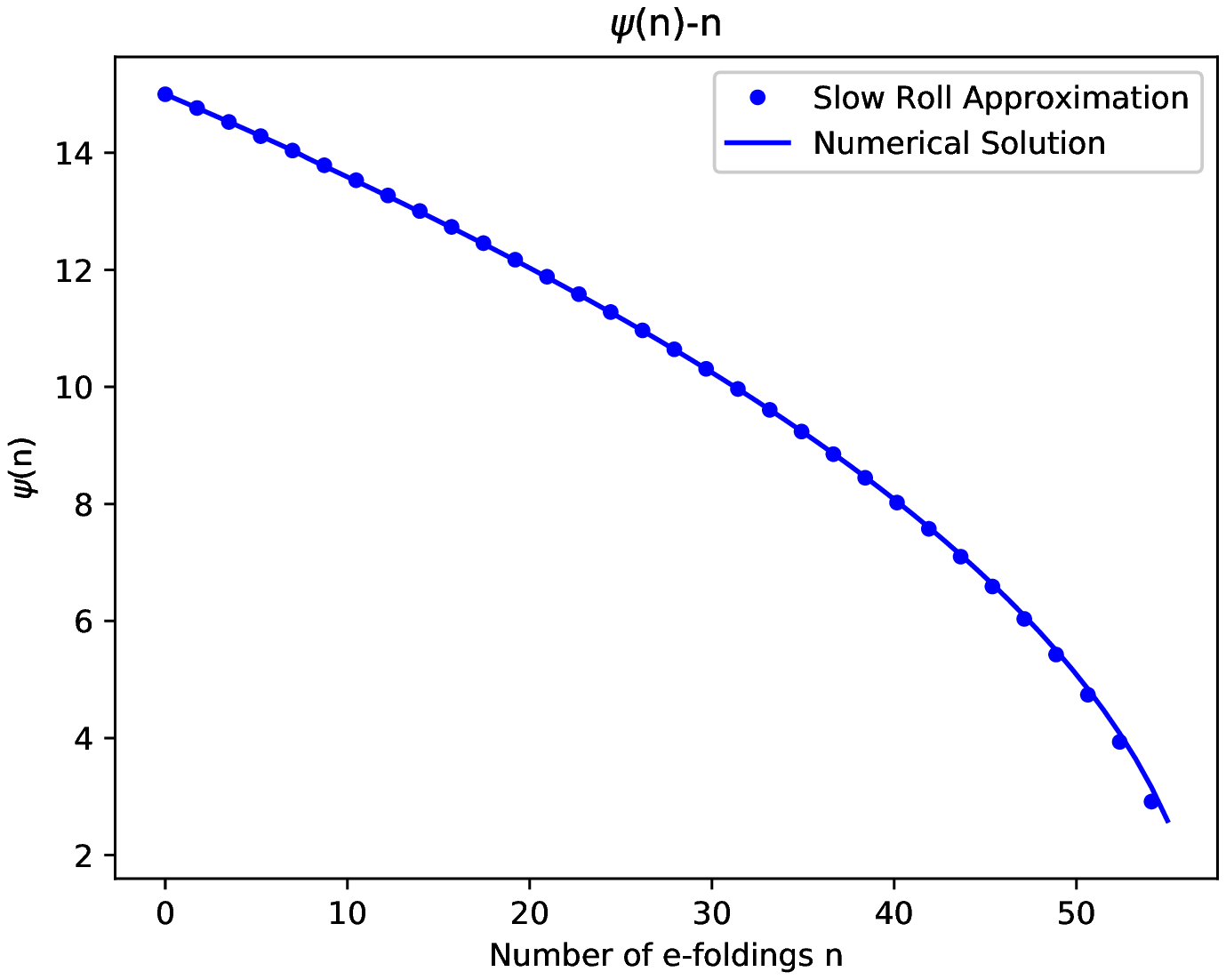}
\hspace{-0.5cm}
\includegraphics[width=4.75cm,height=4.75cm]{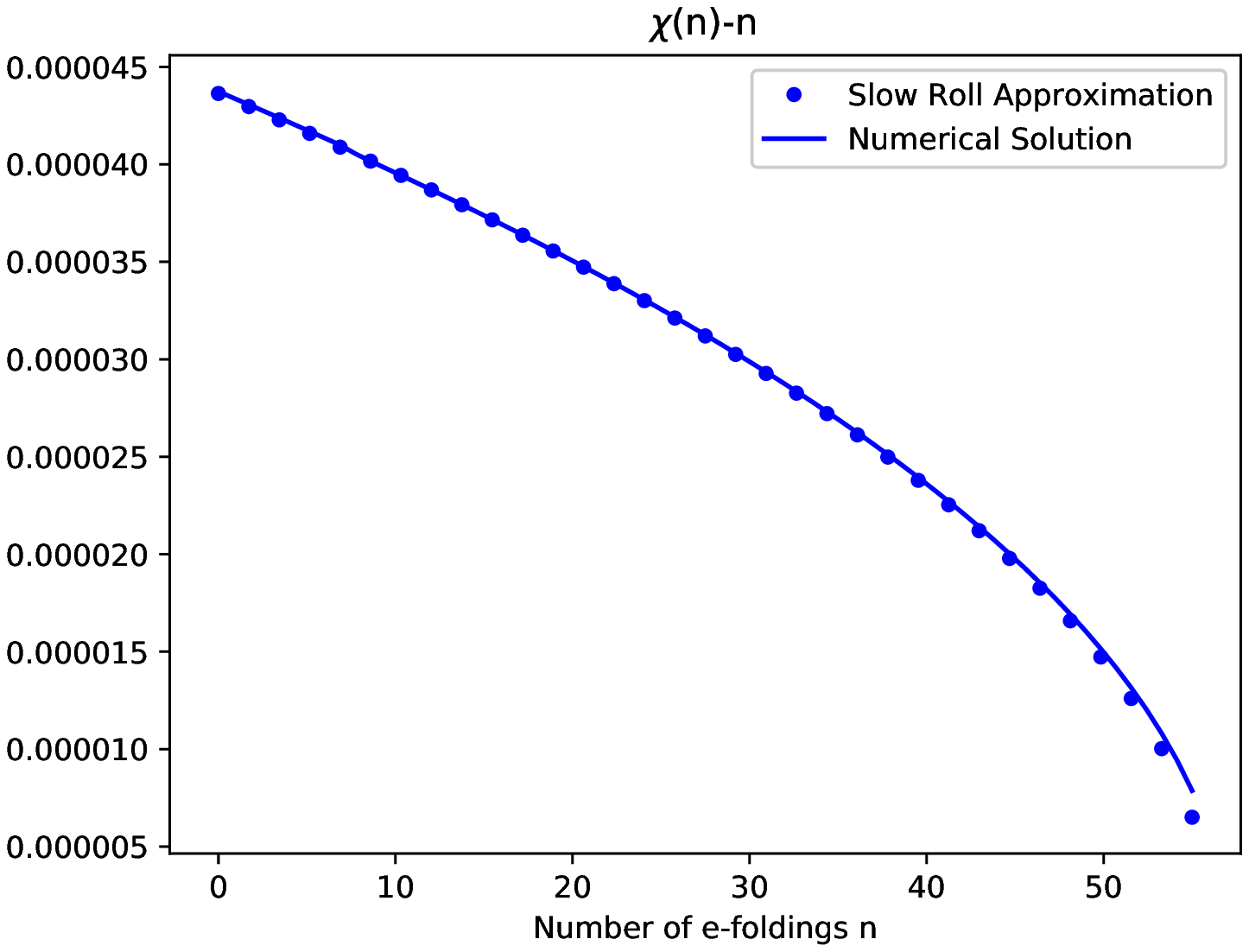}
\hspace{-0.5cm}
\includegraphics[width=4.75cm,height=4.75cm]{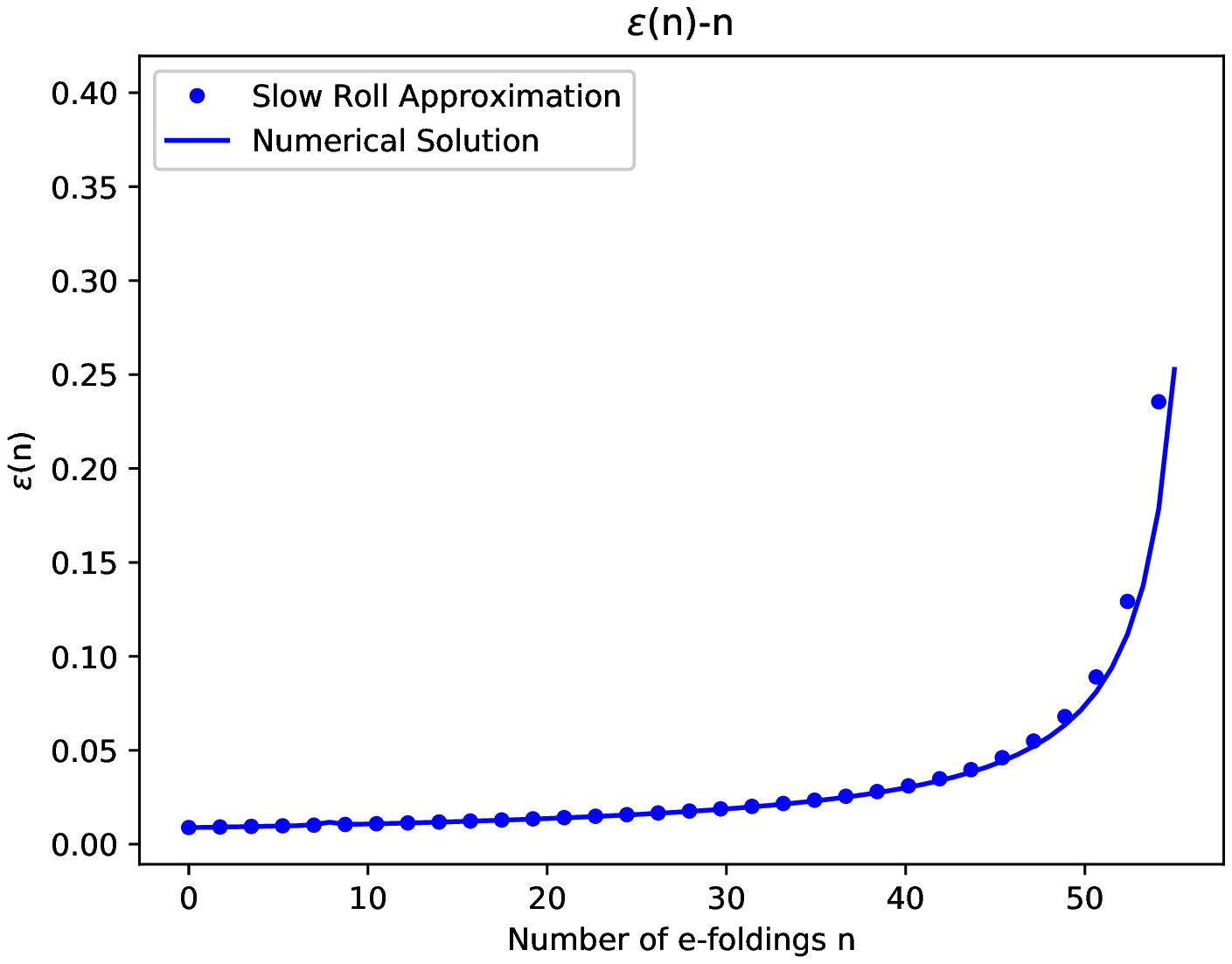}
\caption{\footnotesize{These graphs show the quantities $\psi(n)$, $\chi(n)$ and $\epsilon(n)$ 
for $U = \frac12 c^2 \psi^2$ (with $c = 7.126 \times 10^{-6}$), starting from $\psi_0 = 15$ and
$\psi_0' = -\frac2{\psi_0}$. In each case numerical results for (\ref{Infl4}-\ref{Infl5}) are
plotted as solid lines with the slow roll approximations (\ref{slowroll}) overlaid in large
dots.}}
\label{geometry}
\end{figure}

The numerical evolution of the mode functions requires specialization to a
particular model of inflation. For simplicity we have chosen the quadratic potential 
$U(\psi) = \frac12 c^2 \psi^2$, even though it is no longer consistent with the data. 
Producing the correct amplitude of scalar density perturbations in this model requires 
$c \simeq 7.1 \times 10^{-6}$ \cite{Aghanim:2018eyx}, and we can get about 50 
e-foldings of inflation with initial value $\psi_0 = 15$. The slow roll approximation
for this model gives,
\begin{equation}
\psi(n) \simeq \sqrt{\psi_0^2 \!-\! 4n} \quad , \quad \chi(n) \simeq \frac{c}{\sqrt{6}} 
\sqrt{\psi_0^2 \!-\! 4n} \quad , \quad \epsilon(n) \simeq \frac{2}{\psi_0^2 \!-\! 4 n}
\; . \label{slowroll}
\end{equation}
Figure~\ref{geometry} compares the slow roll approximations (\ref{slowroll}) with exact
numerical evolution of (\ref{Infl4}-\ref{Infl5}).  
\noindent Because there is no perceptible difference we evolved the mode functions using
the slow roll expressions (\ref{slowroll}).

We scale out the dimensions of the wave number and the $\phi$ mass, as well as taking the
logarithm of the norm-squared of the mode functions,
\begin{equation}
\kappa \equiv \sqrt{8 \pi G} \, k \quad , \quad \mu \equiv \sqrt{8\pi G} \, M \quad , \quad
\mathcal{M}(n,\kappa,\mu) \equiv \ln\Biggl[ \frac{\vert u(t,k,M)\vert^2}{\sqrt{8\pi G}}
\Biggr] \; . \label{dimeless2}
\end{equation}
After some manipulations (for all the details when $M = 0$, see \cite{Romania:2012tb,
Brooker:2015iya}) the mode equation and the Wronskian (\ref{ueqns}) can be combined to
give a single, nonlinear equation for $\mathcal{M}(n,\kappa,\mu)$,
\begin{equation}
\mathcal{M}'' + \frac12 {\mathcal{M}'}^2 + (3 \!-\! \epsilon) \mathcal{M}' 
+ \frac{2 \kappa^2 e^{-2 n}}{\chi^2} + \frac{2 \mu^2}{\chi^2} - 
\frac{ \exp[-6n - 2 \mathcal{M}]}{2 \chi^2} = 0 \; . \label{Meqn}
\end{equation}
In the far ultraviolet the physical wave number is much large than either the Hubble
parameter or the mass and the mode function has the WKB form,
\begin{equation}
\frac{k}{a} \gg \Bigl\{ H, M \Bigr\} \quad \Longrightarrow \quad u(t,k,M) \simeq 
\frac1{\sqrt{2 k a^2(t)}} \, \exp\Biggl[ -ik \!\! \int_{t_i}^{t} \!\! \frac{dt'}{a(t')} 
\Biggr] \; . \label{WKB}
\end{equation}
The WKB form (\ref{WKB}) implies initial conditions for $\mathcal{M}(n,\kappa,\mu)
\simeq -\ln(2\kappa) - 2n$,
\begin{equation}
\mathcal{M}(0,\kappa,\mu) = \ln\Bigl[ \frac1{2\kappa} \Bigr] \qquad , \qquad
\mathcal{M}'(0,\kappa,\mu) = -2 \; . \label{initial}
\end{equation}  

\subsection{Phases of $\mathcal{M}(n,\kappa,\mu)$}

\begin{figure}[H]
\includegraphics[width=6.0cm,height=4.0cm]{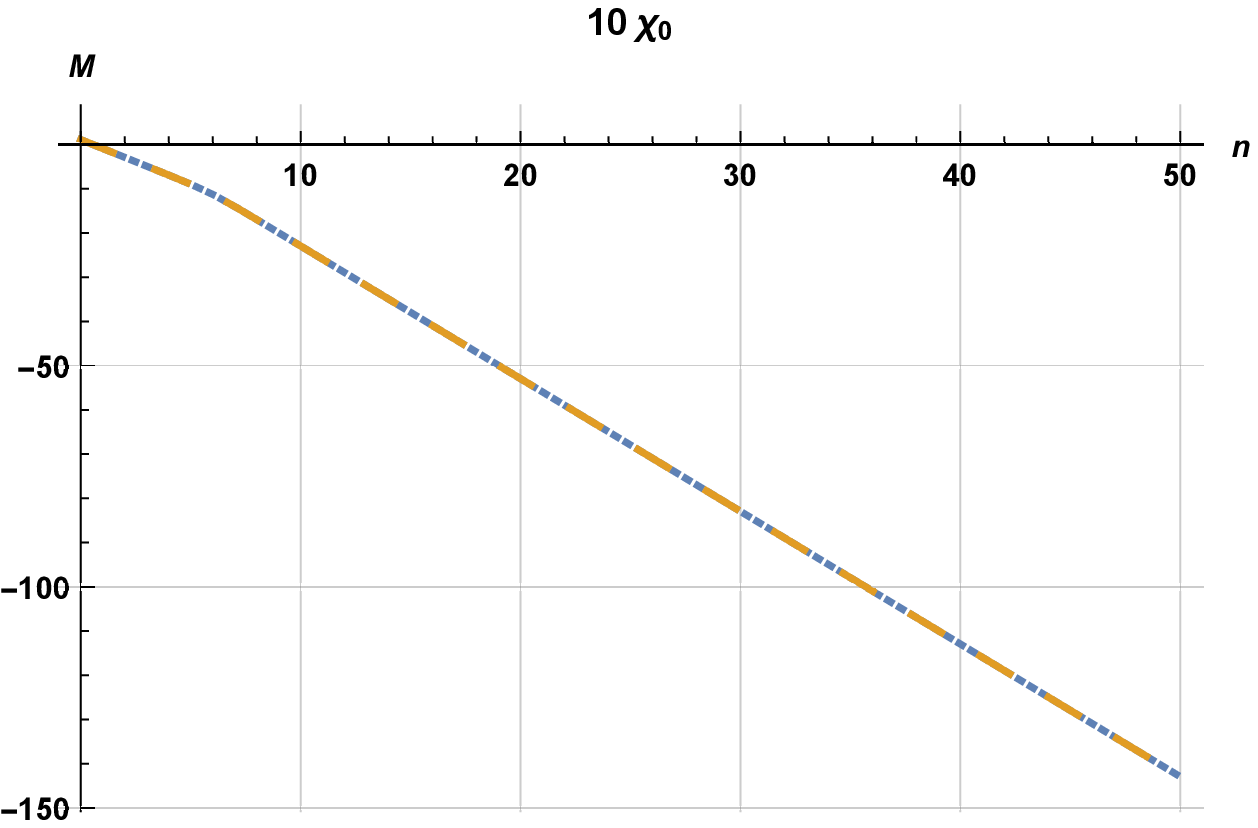}
\hspace{.5cm}
\includegraphics[width=6.0cm,height=4.0cm]{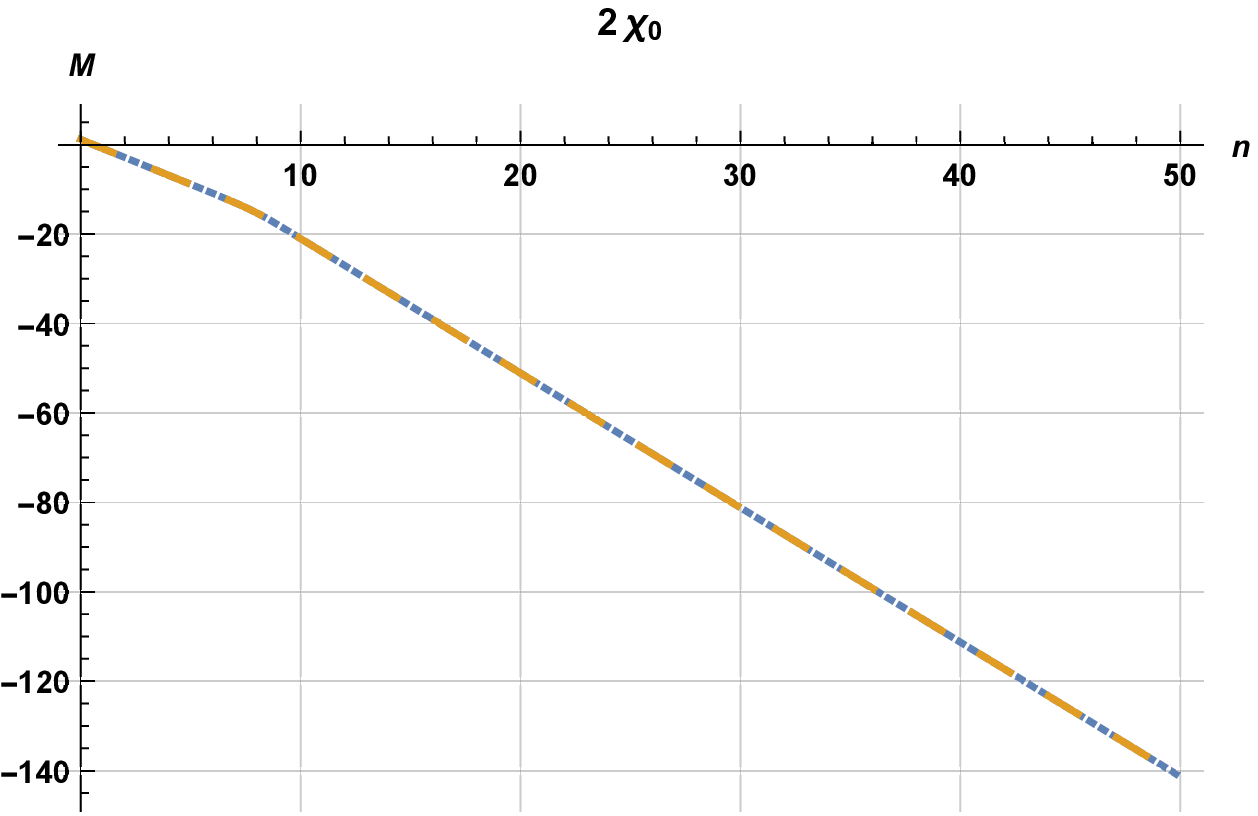}
\caption{\footnotesize{Both plots concern $\kappa = 3800 \chi_0$ (with $n_{\kappa} \simeq 8.32$) and 
$\omega^2(n,\mu) < 0$ throughout inflation. The left hand graph compares the numerical result for 
$\mathcal{M}(n,\kappa,\mu)$ (in blue dots) with the approximation $\mathcal{M}_1(n,\kappa,\mu)$ 
(in long yellow dashes) given in expression (\ref{M1def}) for $\mu = 10 \chi_0$. The right hand
makes the same comparison for $\mu = 2 \chi_0$.}}
\label{mu10and2}
\end{figure}

The amplitude function $\mathcal{M}(n,\kappa,\mu)$ exhibits a number of different
behaviors which depend upon how the physical wave number $\kappa e^{-n}$ and the mass 
$\mu$ relate to the Hubble parameter $\chi(n)$. Two key wave numbers $n_{\kappa}$ and 
$n_{\mu}$ are defined by the relations,
\begin{eqnarray}
{\rm Horizon\ Crossing} & \Longrightarrow & \kappa \equiv e^{n_{\kappa}} \chi(n_{\kappa}) 
\; , \label{nkappa} \\
{\rm Mass\ Domination} & \Longrightarrow & \mu \equiv \frac32 \chi(n_{\mu}) \; . 
\label{nmu}
\end{eqnarray}
For some parameter choices Horizon Crossing and/or Mass Domination may occur before the 
start of inflation, or after its end (at $n_e$), and it may of course be that $n_{\mu} < 
n_{\kappa}$. However, under the ``normal'' assumption that $0 < n_{\kappa} < n_{\mu} < n_{e}$
we distinguish three phases of evolution:
\begin{enumerate}
\item{{\bf Ultraviolet}, for $0 \leq n \ltwid n_{\kappa} + 4$;}
\item{{\bf Steady Decline}, for $n_{\kappa} + 4 \ltwid n \leq n_{\mu}$; and}
\item{{\bf Oscillatory Decline}, for $n_{\mu} \leq n \leq n_{e}$.}
\end{enumerate}
During the first phase $\mathcal{M}(n,\kappa,\mu)$ is well approximated by,
\begin{equation}
\mathcal{M}_1(n,\kappa,\mu) \equiv \ln\Biggl[ \frac1{2 \kappa e^{(D-2) n}} \!\times\!
\frac{\pi}{2} \!\times\! z(n,\kappa) \!\times\! \Bigl\vert H^{(1)}_{\nu(n,\mu)}\Bigl(
z(n,\kappa) \Bigr) \Bigr\vert^2 \Biggr] \; , \label{M1def}
\end{equation}
where the argument $z(n,\kappa)$ and the index $\nu(n,\mu)$ are,
\begin{equation}
z(n,\kappa) \equiv \frac{\kappa e^{-n}}{[1 \!-\! \epsilon(n)] \chi(n)} \;\; , \;\;
\nu^2(n,\mu) \equiv \frac14 \Biggl[ \frac{D \!-\! 1 \!-\! \epsilon(n)}{1 \!-\! 
\epsilon(n)} \Biggr]^2 - \Biggl[ \frac{\mu}{[1 \!-\! \epsilon(n)] \chi(n)} \Biggr]^2 .
\label{znudef}
\end{equation}
We shall always choose $\kappa$ so as to make $n_{\kappa} > 0$, but we explore various
choices of $\mu$. For relatively large masses, such as the cases of $\mu = 10 \chi(0)$ 
and $\mu = 2 \chi(0)$ which are shown in Figure~\ref{mu10and2}, the system is mass 
dominated ($n_{\mu} < 0$) from the beginning of inflation and expression (\ref{M1def}) 
is an excellent approximation throughout inflation.

\begin{figure}[H]
\includegraphics[width=6.0cm,height=4.0cm]{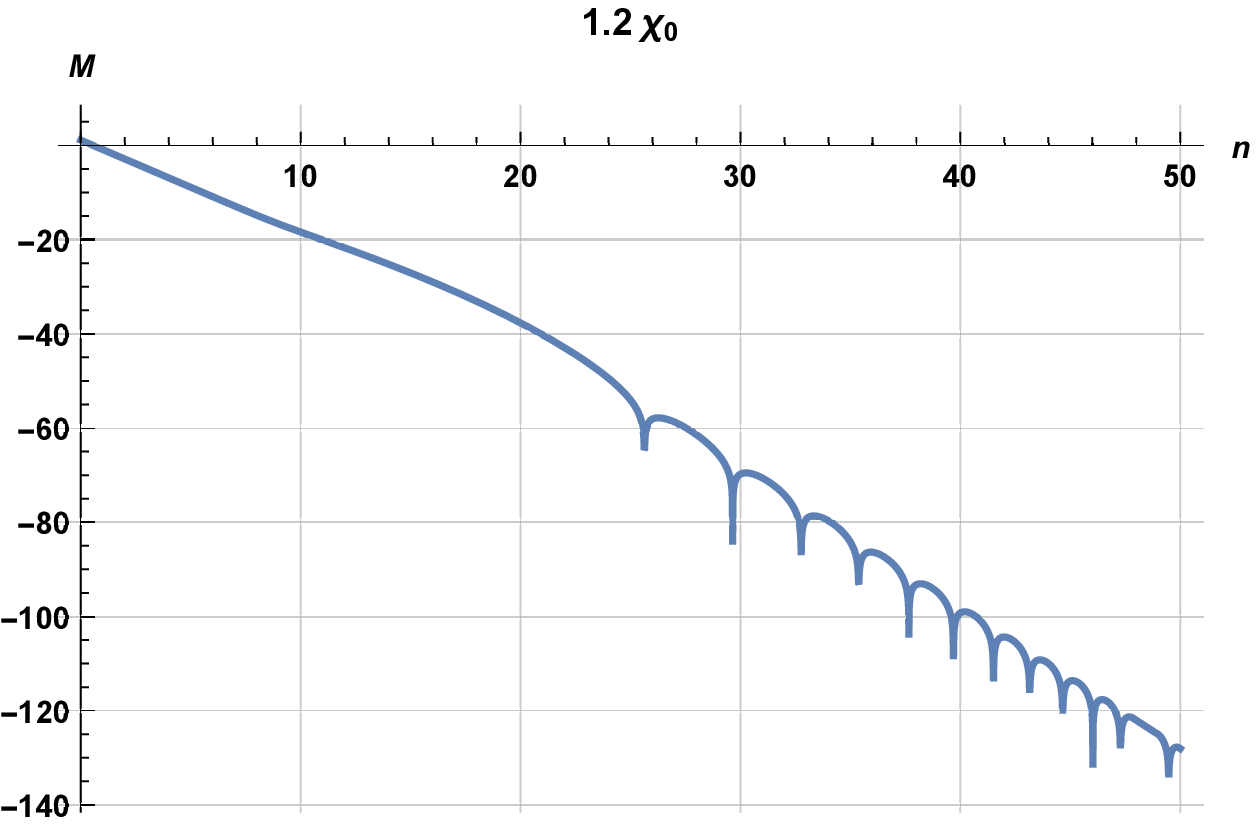}
\hspace{.5cm}
\includegraphics[width=6.0cm,height=4.0cm]{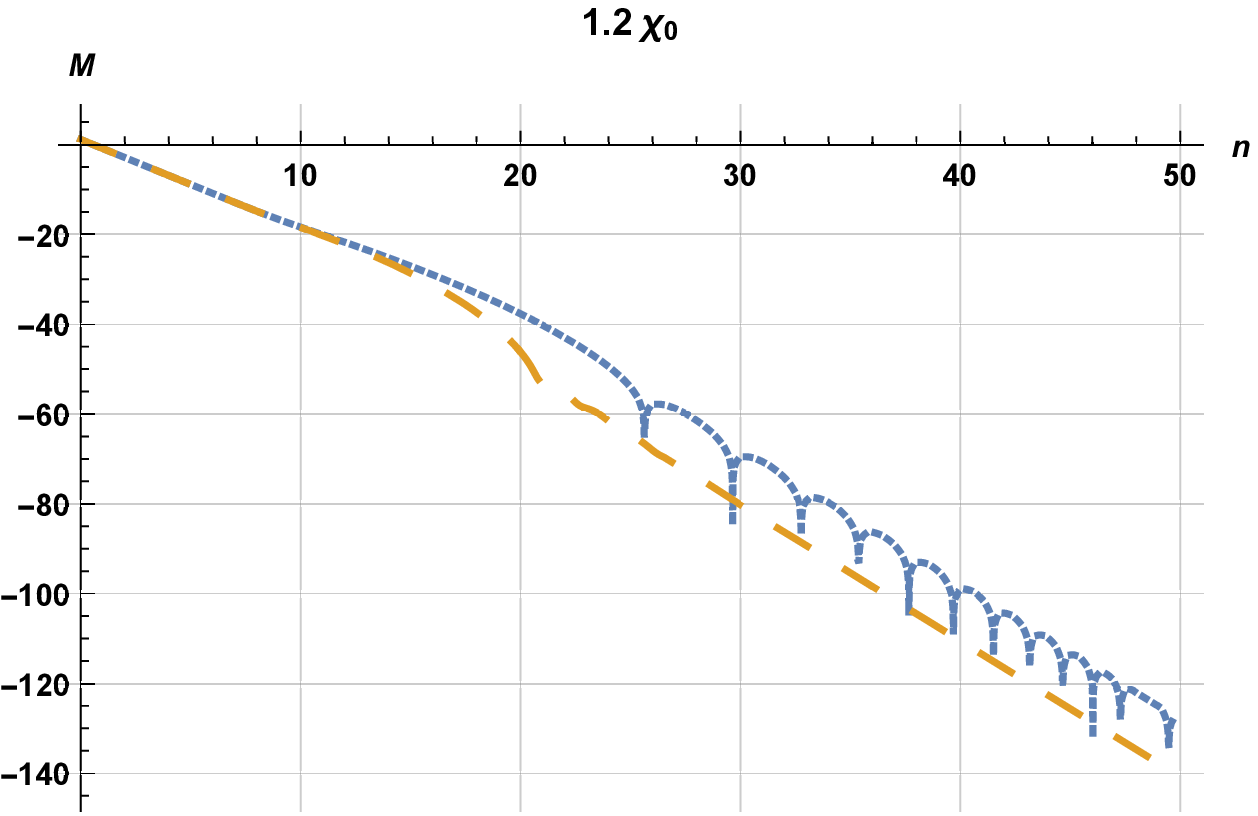}
\caption{\footnotesize{Both plots concern $\kappa = 3800 \chi_0$ (with $n_{\kappa} \simeq 8.32$) and 
$\mu = 1.2 \chi_0$ (with $n_{\mu} \simeq 20.25$). The left hand graph shows the numerical 
result for $\mathcal{M}(n,\kappa,\mu)$. The right hand graph compares this (in blue dots) 
with the approximation $\mathcal{M}_1(n,\kappa,\mu)$ (in long yellow dashes) given in 
expression (\ref{M1def}).}}
\label{mu12A}
\end{figure}

\begin{figure}[H]
\includegraphics[width=6.0cm,height=4.0cm]{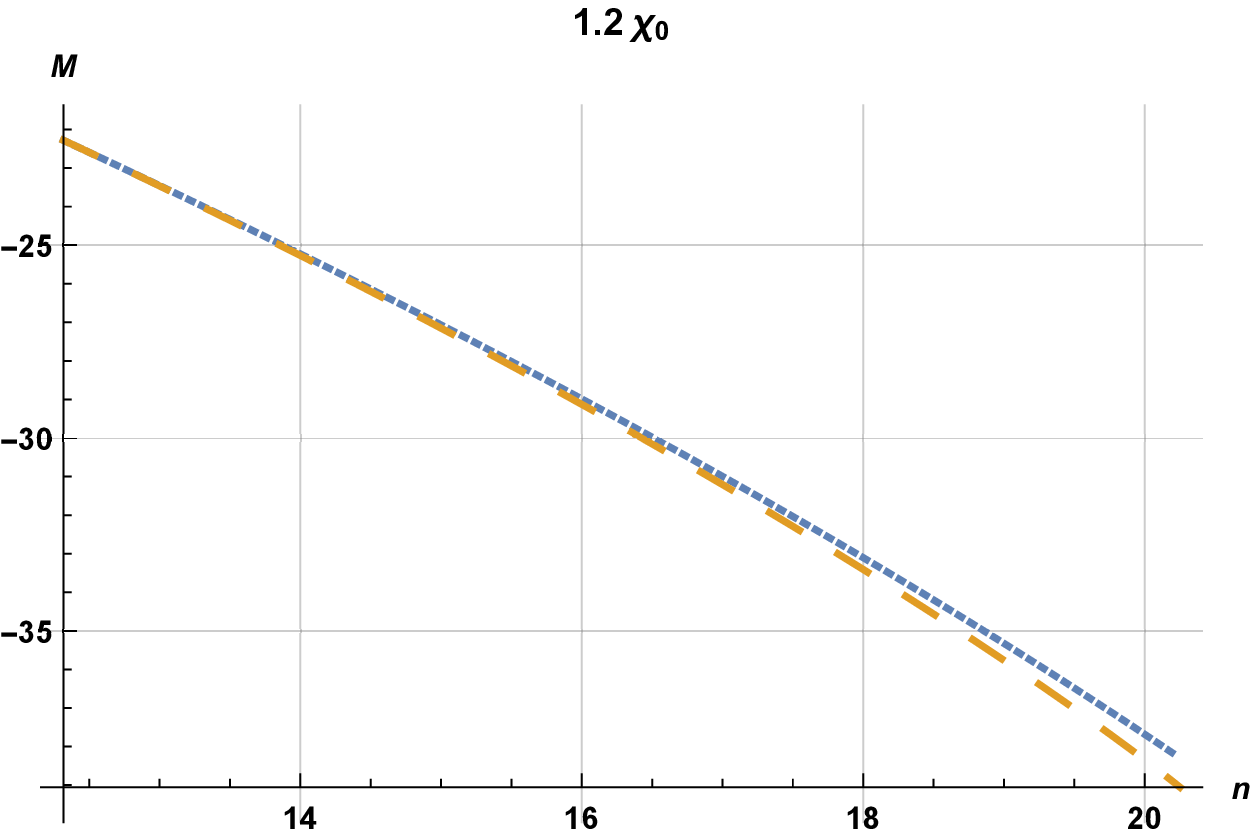}
\hspace{.5cm}
\includegraphics[width=6.0cm,height=4.0cm]{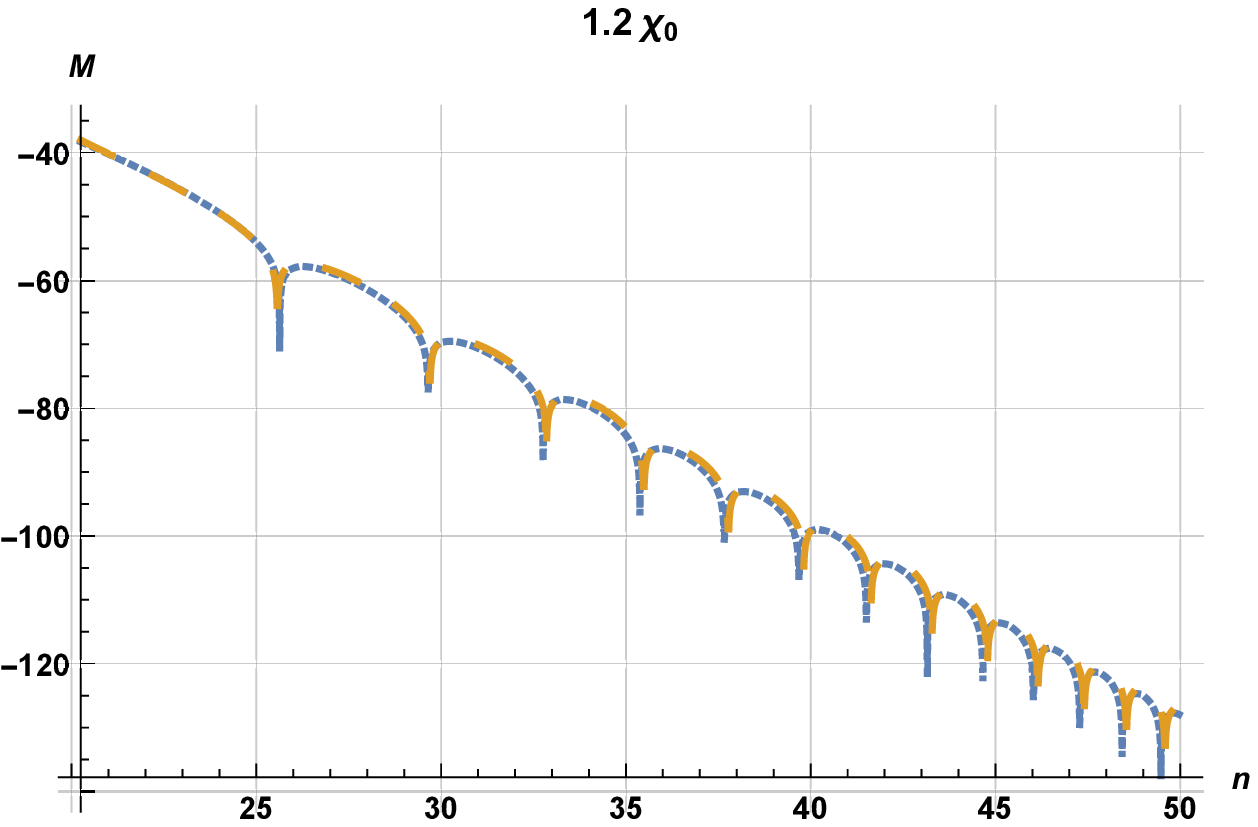}
\caption{\footnotesize{Like the previous figure, these plots deal concern $\kappa = 3800 \chi_0$ 
(with $n_{\kappa} \simeq 8.32$) and $\mu = 1.2 \chi_0$ (with $n_{\mu} \simeq 20.25$).
The left hand graph compares the numerical result for $\mathcal{M}(n,\kappa,\mu)$ 
(in blue dots) with the ``Steady Decline'' approximation $M_2(n,\kappa,\mu)$ (in long 
yellow dashes) given in expression (\ref{M2def}). The right hand graph compares 
$\mathcal{M}(n,\kappa,\mu)$ (in blue dots) with the ``Oscillatory Decline'' 
approximation $\mathcal{M}_3(n,\kappa,\mu)$ (in long yellow dashes) given in 
expression (\ref{M3def}).}}
\label{mu12B}
\end{figure}

For smaller values of $\mu$ the onset of mass domination occurs after horizon 
crossing and the ultraviolet approximation (\ref{M1def}) breaks down. This is shown
for the case of $\mu = 1.2 \chi(0)$ in Figure~\ref{mu12A}.
\noindent In these cases it is useful to define differential frequency functions which 
are real on either side of mass domination,  
\begin{equation}
\omega^2(n,\mu) \equiv \frac{9}{4} - \frac{\mu^2}{\chi^2(n)} \qquad , \qquad
\Omega^2(n,\mu) \equiv \frac{\mu^2}{\chi^2(n)} - \frac{9}{4} \; . \label{freqs}
\end{equation}
If we define $n_2 \equiv n_{\kappa} + 4$, a good approximation for the second phase is,
\begin{eqnarray}
\lefteqn{\mathcal{M}_2(n,\kappa,\mu) = \mathcal{M}_2 - 3 (n \!-\! n_2) } \nonumber \\
& & \hspace{0cm} + 2 \ln\Biggl\{\cosh\Biggl[ \int_{n_2}^{n} \!\!\! dn' \, \omega(n',\mu) 
\Biggr] + \Bigl(\frac{\mathcal{M}_2'\!+\! 3}{2 \omega(n_2,\mu)}\Bigr) \sinh\Biggl[ 
\int_{n_2}^{n} \!\!\! dn' \, \omega(n',\mu)\Biggr]\Biggr\} . \qquad \label{M2def}
\end{eqnarray}
Defining $n_3 \equiv n_{\mu} + 4$ gives a good approximation for the third phase,
\begin{eqnarray}
\lefteqn{\mathcal{M}_3(n,\kappa,\mu) = \mathcal{M}_3 - 3 (n \!-\! n_3) } \nonumber \\
& & \hspace{0cm} + 2 \ln\Biggl\{\Biggl\vert \cos\Biggl[ \int_{n_3}^{n} \!\!\! dn' \, 
\Omega(n',\mu) \Biggr] + \Bigl(\frac{\mathcal{M}_3'\!+\! 3}{2 \Omega(n_3,\mu)}\Bigr) 
\sin\Biggl[ \int_{n_3}^{n} \!\!\! dn' \, \Omega(n',\mu)\Biggr]\Biggr\vert\Biggr\} . 
\qquad \label{M3def}
\end{eqnarray}
Figure~\ref{mu12B} demonstrates the validity of these approximations for $\mu = 1.2 
\chi(0)$.

\begin{figure}[H]
\includegraphics[width=6.0cm,height=4.0cm]{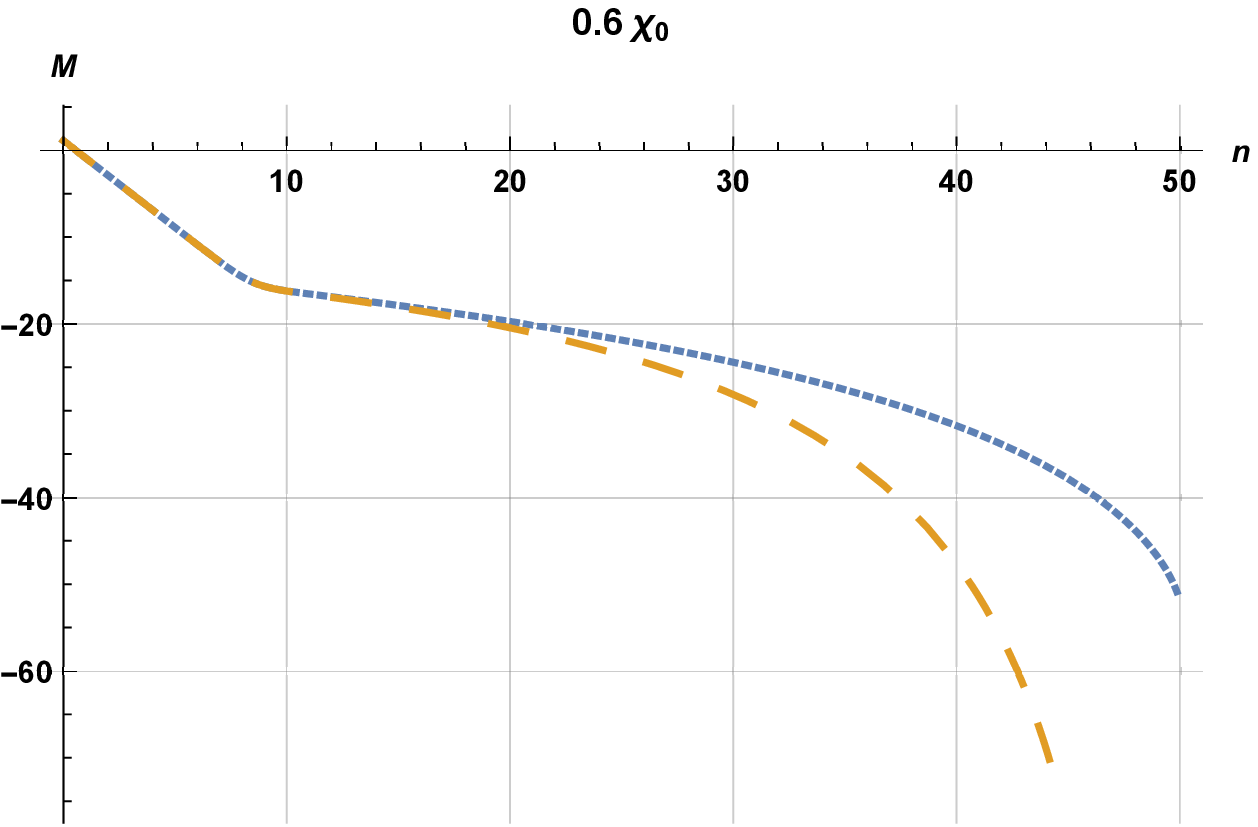}
\hspace{.5cm}
\includegraphics[width=6.0cm,height=4.0cm]{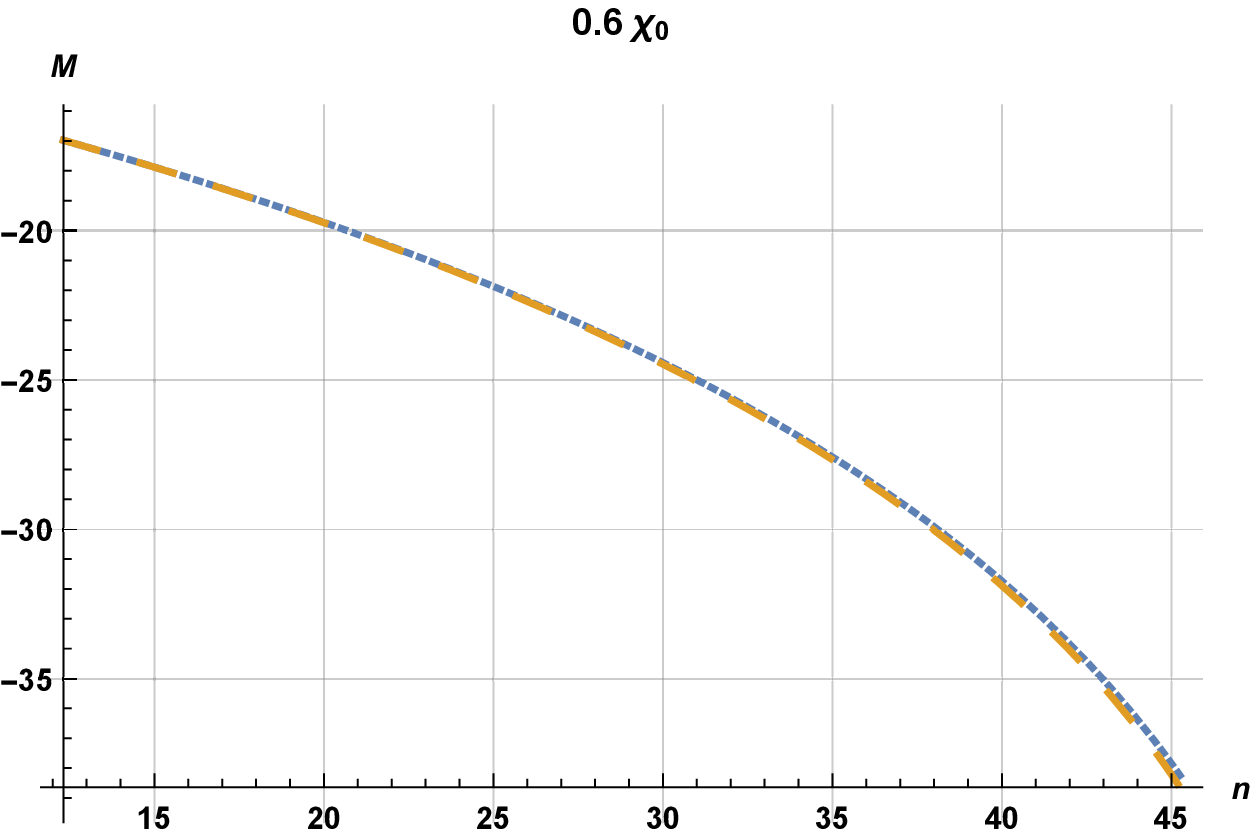}
\caption{\footnotesize{Both plots concern $\kappa = 3800 \chi_0$ (with $n_{\kappa} \simeq 8.32$) and 
$\mu = 0.6 \chi_0$ (with $n_{\mu} \simeq 47.25$). The left hand graph compares  
$\mathcal{M}(n,\kappa,\mu)$ (in blue dots) with the approximation $\mathcal{M}_1(n,\kappa,\mu)$ 
(in long yellow dashes) given in expression (\ref{M1def}). The right hand graph compares 
$\mathcal{M}(n,\kappa,\mu)$ (in blue dots) with the approximation $M_2(n,\kappa,\mu)$ 
(in long yellow dashes) given in expression (\ref{M2def}).}}
\label{mu06}
\end{figure}

\begin{figure}[H]
\includegraphics[width=6.0cm,height=4.0cm]{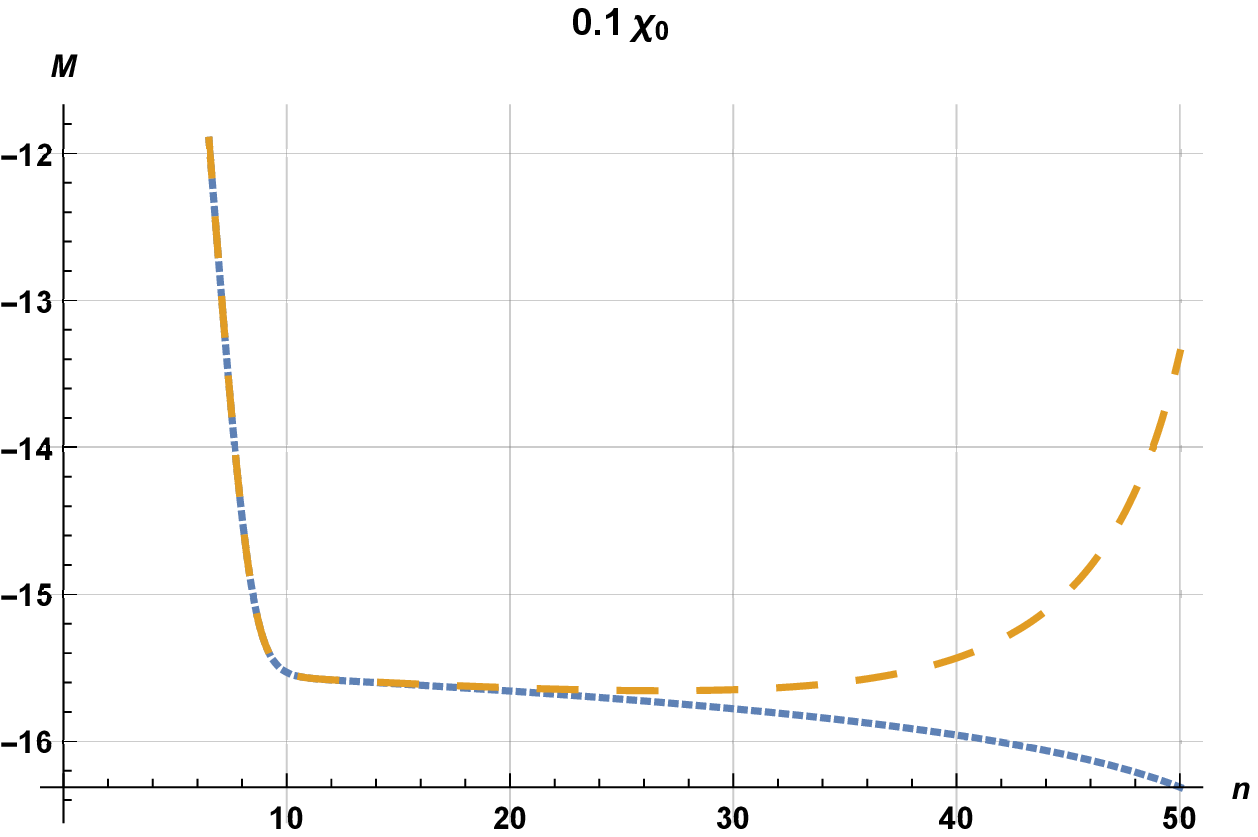}
\hspace{.5cm}
\includegraphics[width=6.0cm,height=4.0cm]{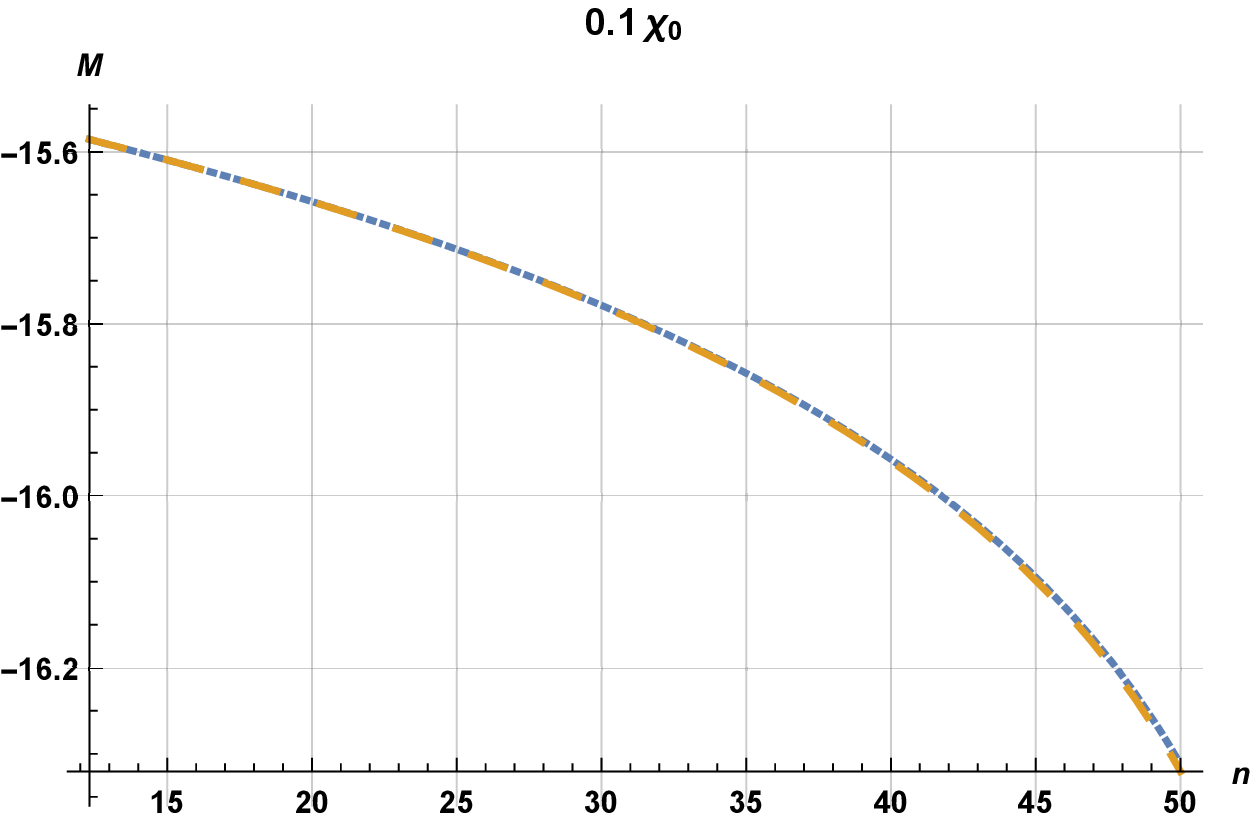}
\caption{\footnotesize{Both plots concern $\kappa = 3800 \chi_0$ (with $n_{\kappa} \simeq 8.32$) and 
$\mu = 0.1 \chi_0$ (with $n_{\mu} > 50$). The left hand graph compares $\mathcal{M}(n,\kappa,\mu)$ 
(in blue dots) with the approximation $\mathcal{M}_1(n,\kappa,\mu)$ (in long yellow dashes) 
given in expression (\ref{M1def}). The right hand graph compares $\mathcal{M}(n,\kappa,\mu)$ 
(in blue dots) with the approximation $M_2(n,\kappa,\mu)$ (in long yellow dashes) given in 
expression (\ref{M2def}).}}
\label{mu01}
\end{figure}

Making $\mu$ smaller postpones the onset of mass domination so late that the third phase
comes near the end of inflation. Figure~\ref{mu06} shows this for $\mu = 0.6 \chi(0)$, which
corresponds to $n_{\mu} \simeq 47.25$.
\noindent For very small values of $\mu$ the onset of mass domination never comes and only the 
first two phases are necessary. Figure~\ref{mu01} shows this for $\mu = 0.1 \chi(0)$, which 
would correspond to $n_{\mu} \simeq 56$ if slow roll inflation persisted that long.

\begin{figure}[H]
\includegraphics[width=4.75cm,height=4.75cm]{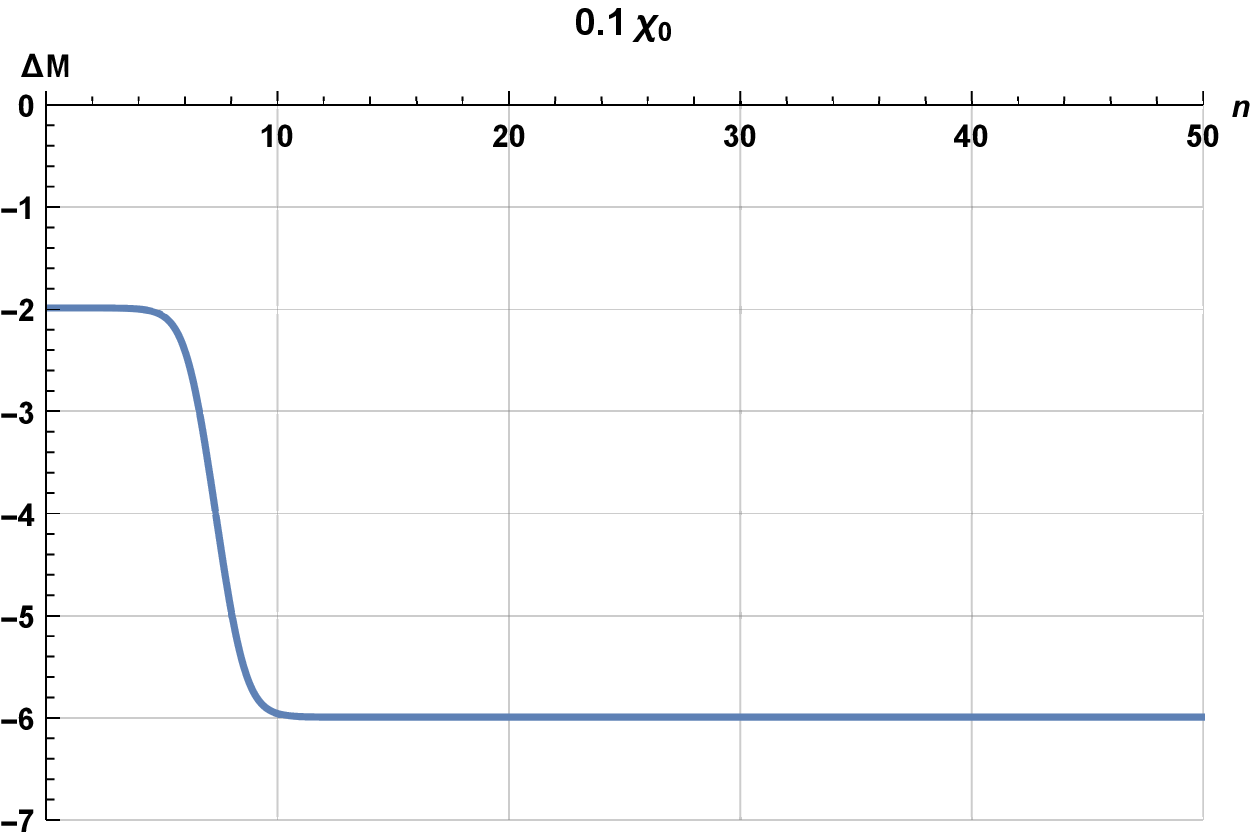}
\hspace{-0.5cm}
\includegraphics[width=4.75cm,height=4.75cm]{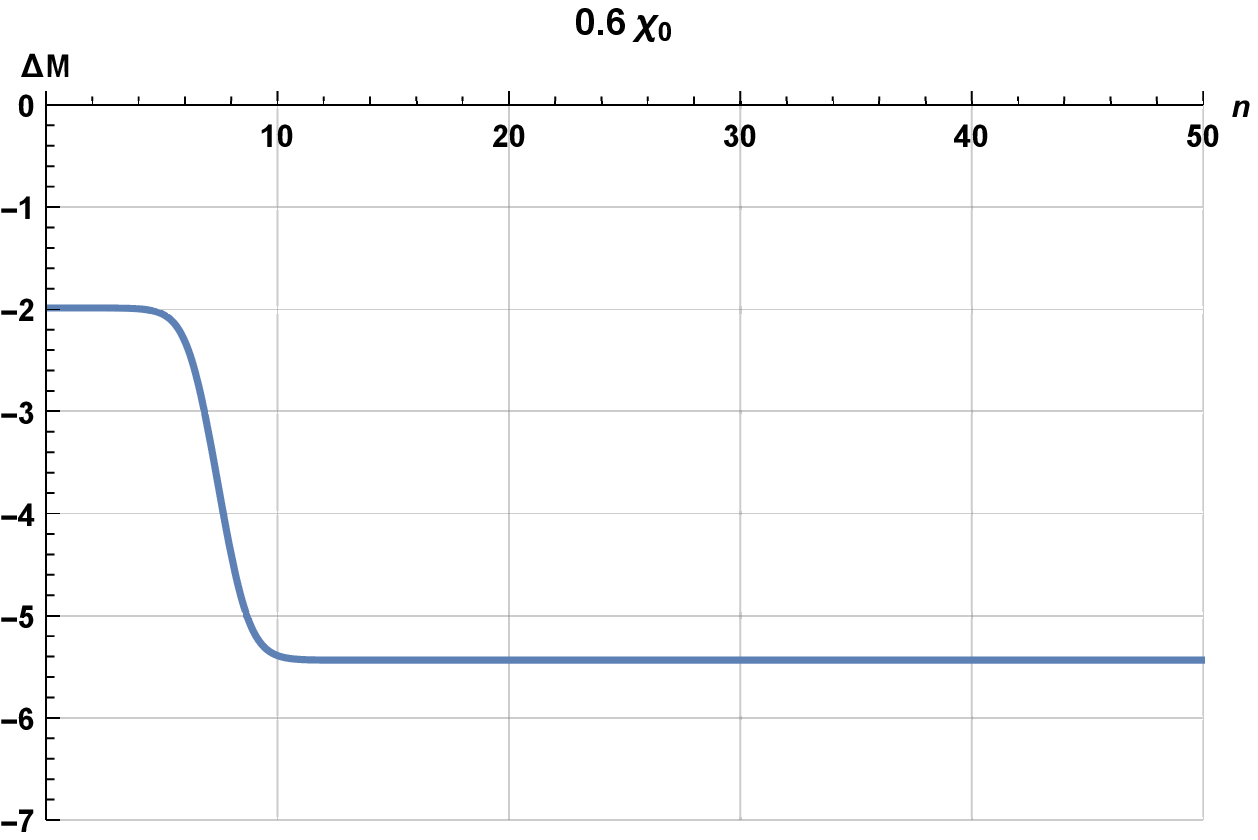}
\hspace{-0.5cm}
\includegraphics[width=4.75cm,height=4.75cm]{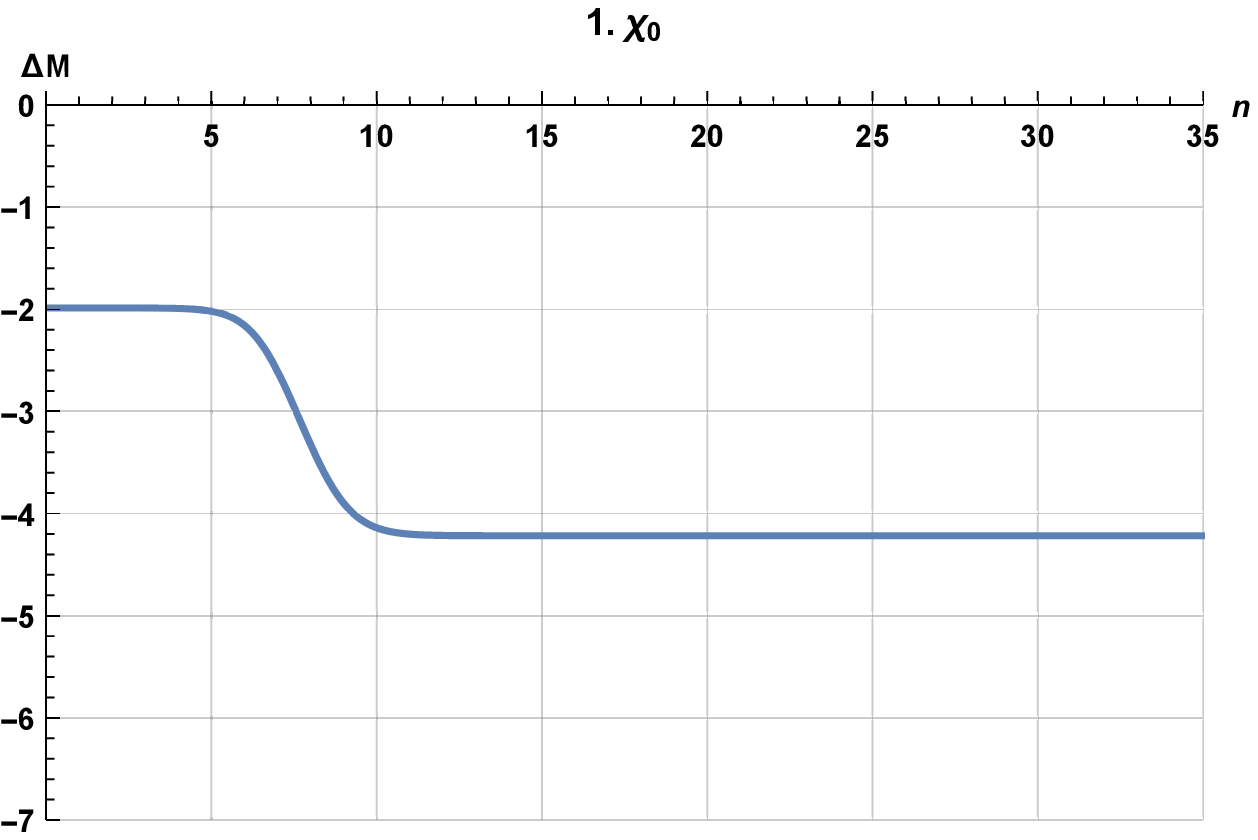}
\caption{\footnotesize{The left hand graph gives the difference of $\mathcal{M}(n,\kappa,\mu)$ between 
$\kappa = 3800 \chi_0$ (with $n_{\kappa} \simeq 8.32$) and $\kappa = 520 \chi_0$ (with 
$n_{\kappa} \simeq 6.31$) for $\mu = 0.1 \chi_0$. The middle and right hand graphs show the 
same difference for the cases of $\mu = 0.6 \chi_0$ and $\mu = \chi_0$, respectively.}}
\label{kappasmall}
\end{figure}

Because the coincidence limit (\ref{Ftrans}) involves an integration over $\kappa = \sqrt{8\pi G} 
\times k$ it is crucial to determine how $\mathcal{M}(n,\kappa,\mu)$ depends on $\kappa$. This
$\kappa$ dependence is manifest in the ultraviolet approximation (\ref{M1def}), but there is no
explicit $\kappa$ dependence in either of the later approximations (\ref{M2def}) and (\ref{M3def}).
However, these approximations do depend on integration constants $\mathcal{M}_{2,3}$ and 
$\mathcal{M}_{2,3}'$ which represent the values of $\mathcal{M}(n,\kappa,\mu)$ and its first
derivative with respect to $n$ at $n_2 = n_{\kappa} + 4$ and $n_3 = n_{\mu} + 4$. It turns out 
that only $\mathcal{M}(n_{2,3},\kappa,\mu)$ depend significantly on $\kappa$. We can see this
numerically by making plots of $\Delta \mathcal{M}(n,\mu) \equiv \mathcal{M}(n,\kappa_1,\mu) -
\mathcal{M}(n,\kappa_2,\mu)$ for fixed wave numbers. Figure~\ref{kappasmall} gives three such 
plots for relatively small values of $\mu$ that would require more than just the ultraviolet 
phase. Because $\Delta \mathcal{M}(n,\mu)$ rapidly freezes in to a constant after horizon 
crossing we see that the two later phases inherit their $\kappa$ dependence from the ultraviolet
phase,
\begin{eqnarray}
\mathcal{M}_2(n,\kappa,\mu) & = & \mathcal{M}_1(n_2,\kappa,\mu) + f_2(n,\mu) \; , 
\label{kappadep2} \\
\mathcal{M}_3(n,\kappa,\mu) & = & \mathcal{M}_1(n_2,\kappa,\mu) + f_3(n,\mu) \; . 
\label{kappadep3}
\end{eqnarray}

\begin{figure}[H]
\includegraphics[width=4.75cm,height=4.75cm]{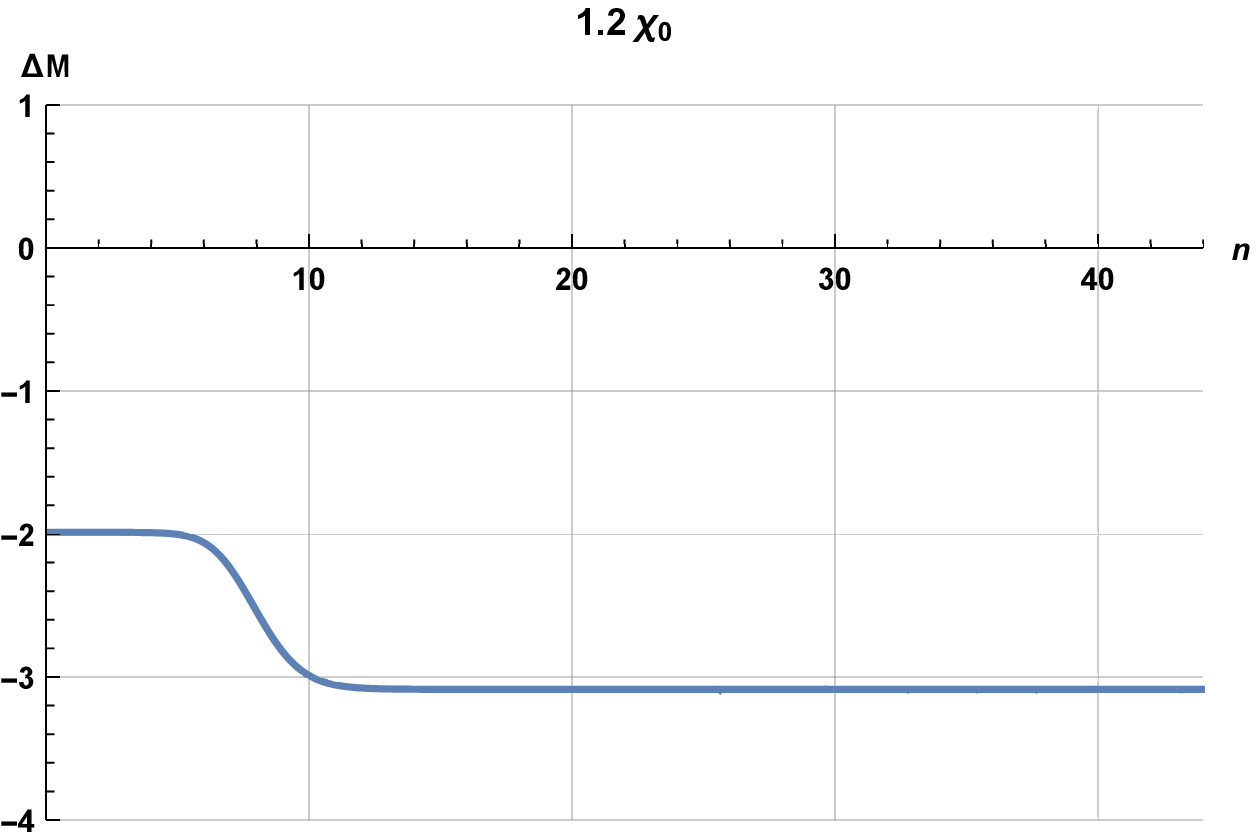}
\hspace{-0.5cm}
\includegraphics[width=4.75cm,height=4.75cm]{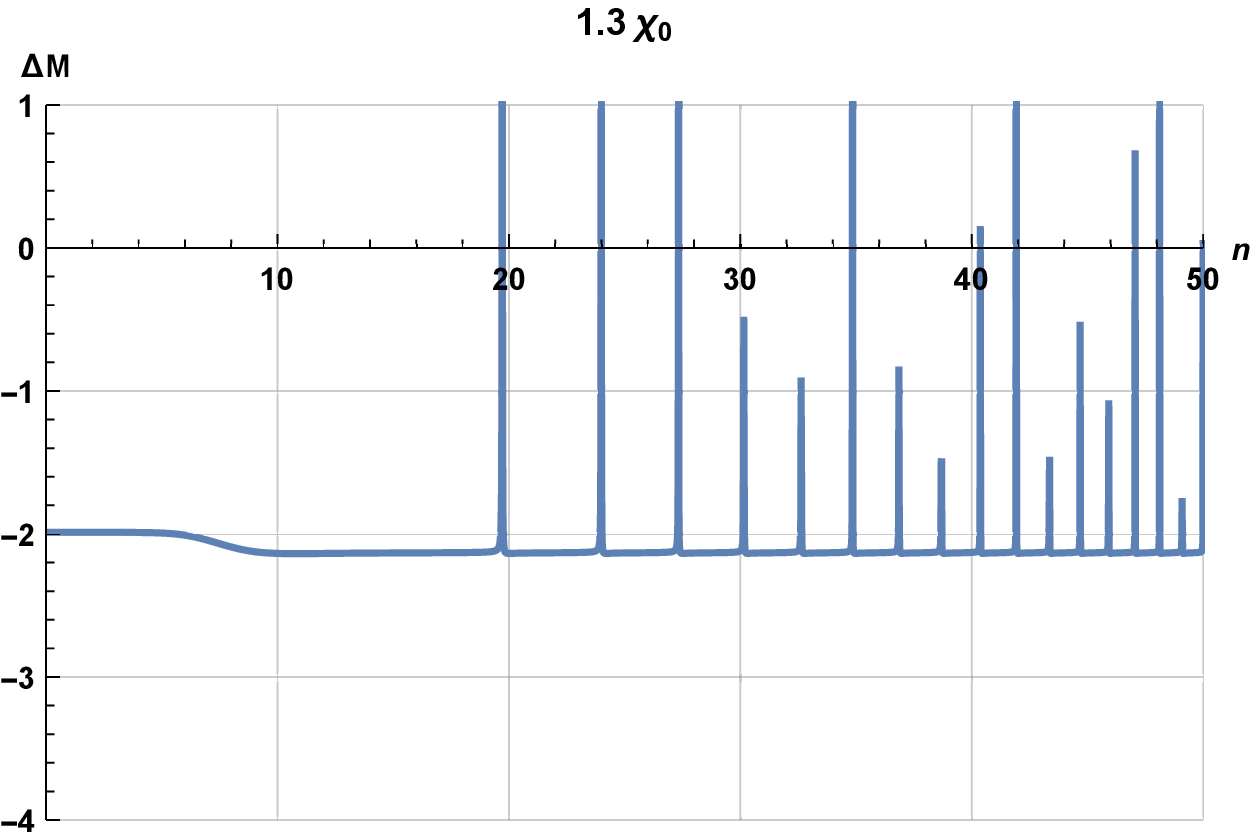}
\hspace{-0.5cm}
\includegraphics[width=4.75cm,height=4.75cm]{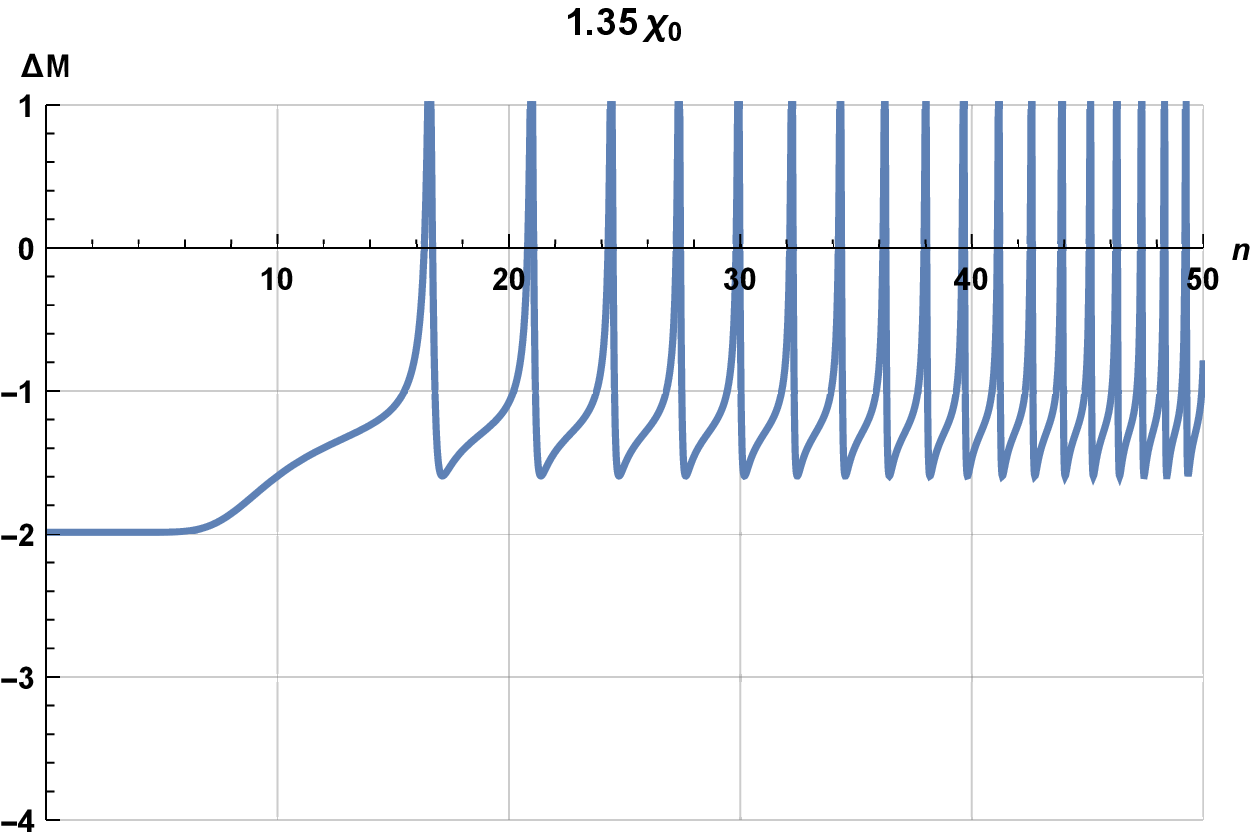}
\caption{\footnotesize{The left hand graph gives the difference of $\mathcal{M}(n,\kappa,\mu)$ between 
$\kappa = 3800 \chi_0$ (with $n_{\kappa} \simeq 8.32$) and $\kappa = 520 \chi_0$ (with 
$n_{\kappa} \simeq 6.31$) for $\mu = 1.2 \chi_0$. The middle and right hand graphs show
the same difference for the cases of $\mu = 1.3 \chi_0$ and $\mu = 1.35 \chi_0$, 
respectively.}}
\label{kappamedium}
\end{figure}

\begin{figure}[H]
\includegraphics[width=4.75cm,height=4.75cm]{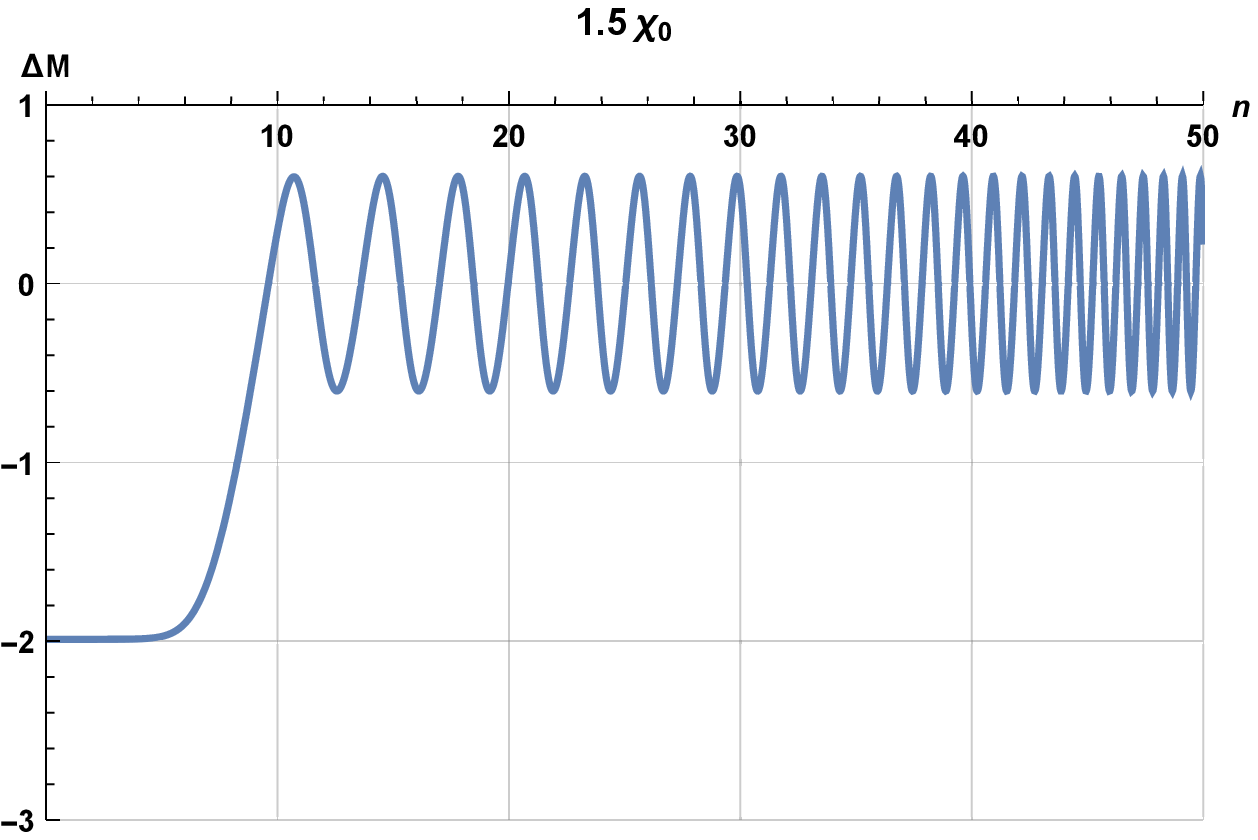}
\hspace{-0.5cm}
\includegraphics[width=4.75cm,height=4.75cm]{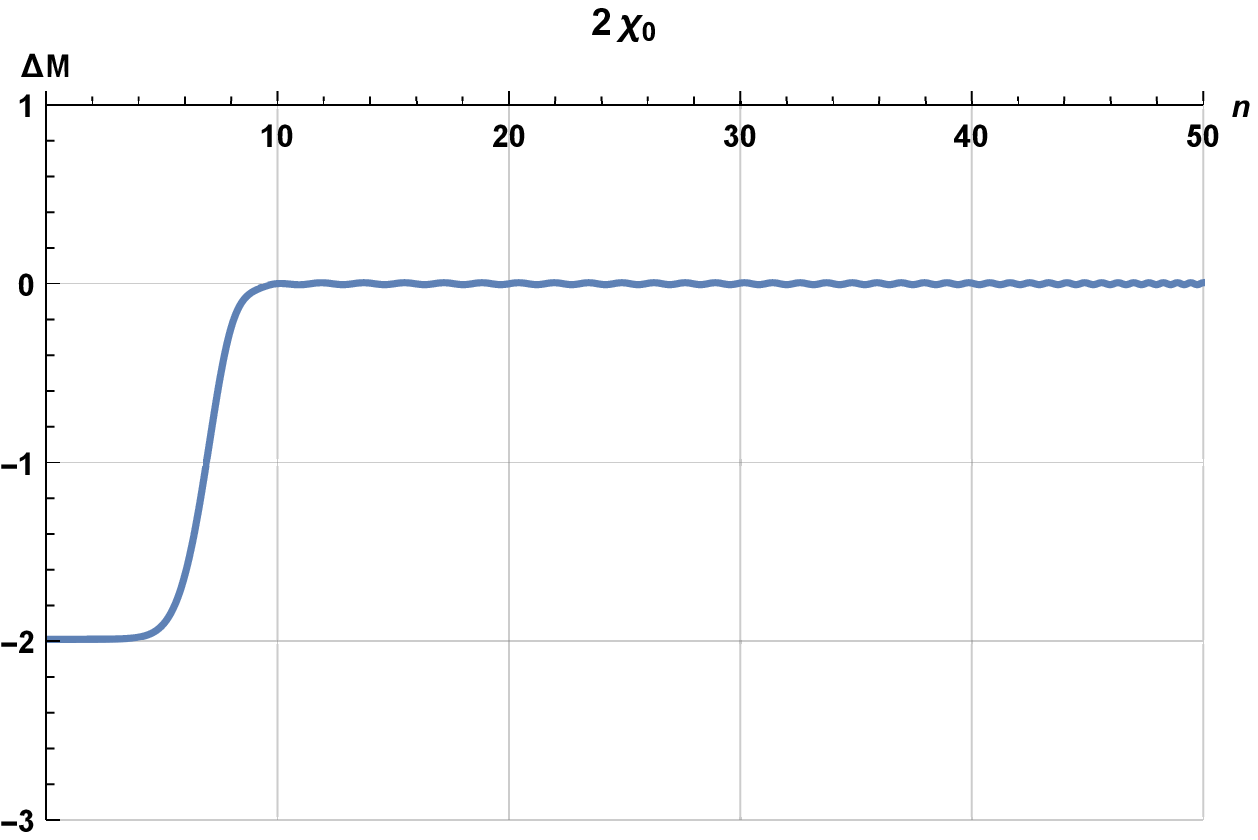}
\hspace{-0.5cm}
\includegraphics[width=4.75cm,height=4.75cm]{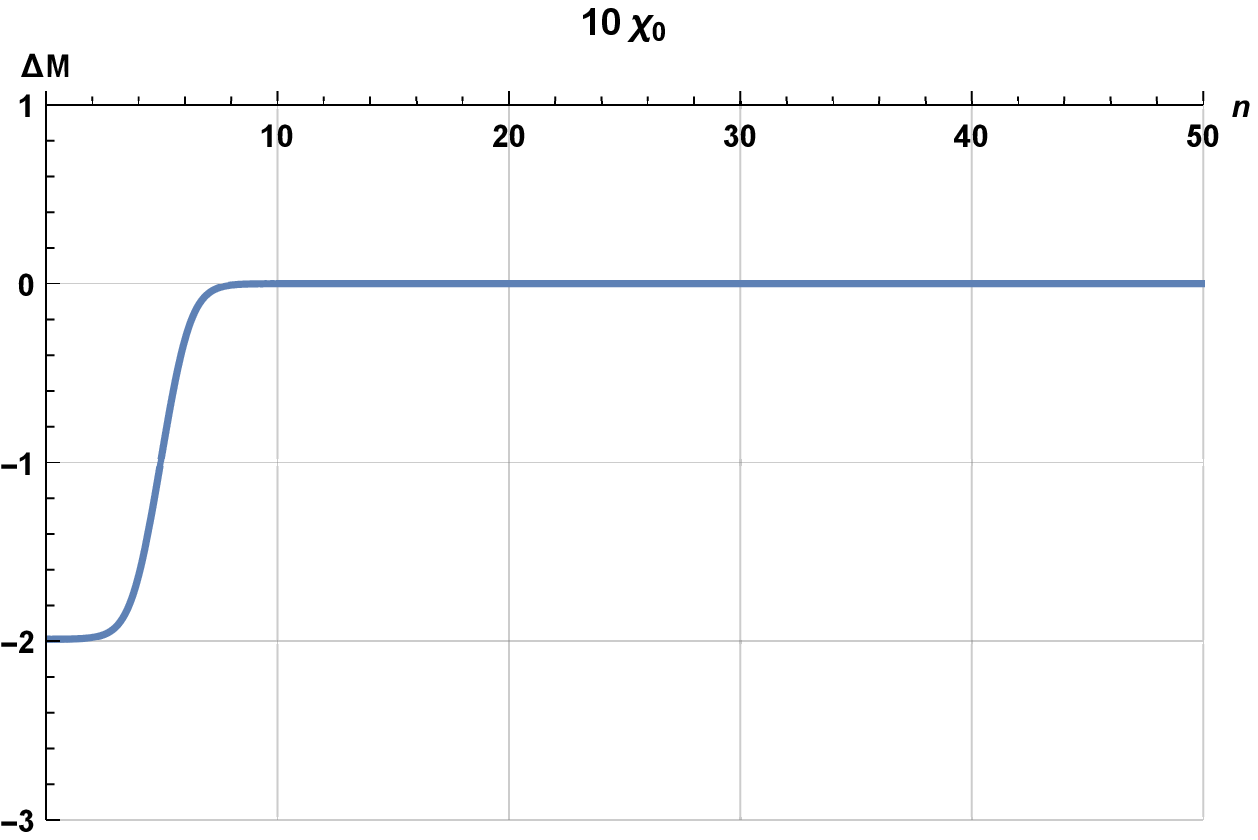}
\caption{\footnotesize{The left hand graph gives the difference of $\mathcal{M}(n,\kappa,\mu)$ between 
$\kappa = 3800 \chi_0$ (horizon crossing at $n_{\kappa} \simeq 8.32$) and $\kappa = 520 \chi_0$ 
(horizon crossing at $n_{\kappa} \simeq 6.31$) for $\mu = 1.5 \chi_0$. For this case the average
post-horizon difference has become centered on zero while the fluctuations have become symmetric
and more frequent yet. The middle and right hand graphs show the same difference for the cases 
of $\mu = 2 \chi_0$ and $\mu = 10 \chi_0$. The difference remains centered on zero with negligibly
small oscillations.}}
\label{kappabig}
\end{figure}

Figure~\ref{kappamedium} gives three plots of $\Delta \mathcal{M}(n,\mu)$ for the intermediate 
values of $\mu$ over which the other phases drop out. 
\noindent All three phases occur for $\mu = 1.2 \chi(0)$, and the difference is constant after 
horizon crossing. As $\mu$ is increased the difference exhibits a complex behavior, but one that
is captured by the ultraviolet approximation (\ref{M1def}). Figure~\ref{kappabig} continues
the progression to even larger values of $\mu$ for which the ultraviolet approximation suffices. 

\subsection{Analytic Derivations}

We had of course to set the spacetime dimension to $D=4$ in order to construct the various 
graphs of the previous sub-section. However, obtaining a reliable ultraviolet form requires
generalizing the evolution equation (\ref{Meqn}) to general $D$,
\begin{equation}
\mathcal{M}'' + \frac12 {\mathcal{M}'}^2 + (D \!-\! 1 \!-\! \epsilon) \mathcal{M}' 
+ \frac{2 \kappa^2 e^{-2 n}}{\chi^2} + \frac{2 \mu^2}{\chi^2} - 
\frac{ \exp[-2 (D - 1) n - 2 \mathcal{M}]}{2 \chi^2} = 0 \; . \label{MDeqn}
\end{equation}
Our initial motivation for the $\mathcal{M}_1(n,\kappa,\mu)$ approximation (\ref{M1def})
was the Hubble effective potential \cite{Janssen:2009pb} that would be relevant for 
constant $\epsilon(n)$ and the case where the mass parameter $\mu = \frac{h}{\sqrt{2}} \psi(n)$ 
is proportional to the Hubble parameter $\chi(n)$. To see that $\mathcal{M}_1(n,\kappa,\mu)$
is generally valid in the ultraviolet, we make the change of variables,
\begin{equation}
\mathcal{M}(n,\kappa,\mu) \equiv \mathcal{M}_1(n,\kappa,\mu) + g(n,\kappa,\mu) 
\equiv -\ln(2 \kappa) - (D \!-\! 2) n + \Delta \mathcal{M}_1 + g \; . \label{gdef}
\end{equation}
Substituting (\ref{gdef}) in (\ref{MDeqn}) gives,
\begin{eqnarray}
\lefteqn{ \hspace{-0.2cm} g'' + \frac{{g'}^2}{2} + \Bigl(1 \!-\! \epsilon \!+\! \Delta \mathcal{M}_1'\Bigr) g'
+ \frac{2 \kappa^2 e^{-2 \Delta \mathcal{M}_1}}{\chi^2 e^{2n}} \Bigl[1 - e^{-2g}\Bigr] = -
\Delta \mathcal{M}_1'' - \frac{{\Delta \mathcal{M}_1'}^2}{2}} \nonumber \\
& & \hspace{0cm} -(1 \!-\! \epsilon) \Delta \mathcal{M}_1' + 2 \Bigl[\Bigl(\frac{D}{2} 
\!-\! 1\Bigr) \Bigl( \frac{D}{2} \!-\! \epsilon \Bigr) - \frac{\mu^2}{\chi^2}\Bigr] + 
\frac{2 \kappa^2}{\chi^2 e^{2n}} \Bigl[ e^{-2 \Delta \mathcal{M}_1} - 1\Bigr] \; . 
\qquad \label{geqn}
\end{eqnarray}
Now make the large $z$ expansion for $\Delta \mathcal{M}_1(n,\kappa,\mu)$,
\begin{equation}
e^{\Delta \mathcal{M}_1} = \frac{\pi}{2} z \Bigl\vert H^{(1)}_{\nu}(z)\Bigr\vert^2 = 1 +
\frac{(\nu^2 \!-\! \frac14)}{2 z^2} + \frac{3 (\nu^2 \!-\! \frac14) (\nu^2 \!-\! \frac94)}{
8 z^4} + O\Bigl(\frac1{z^6}\Bigr) \; . \label{Hankelexp}
\end{equation}
Substituting (\ref{Hankelexp}) into (\ref{geqn}) and solving for $g(n,\kappa,\mu)$ in the 
ultraviolet regime of $\kappa \gg \chi(n) e^{n}$ implies the expansion,
\begin{equation}
g = \frac18 \Biggl[ 2 \epsilon (5 - 3 \epsilon) \frac{\mu^2}{\chi^2}
+ \Bigl( \frac{D}2 - 1\Bigr) \Bigl[(D + 5 - 7 \epsilon) \epsilon' + \epsilon''
\Bigr] \Biggr] \Bigl( \frac{\chi e^n}{\kappa}\Bigr)^4 + O\Biggl( \Bigl( 
\frac{\chi e^n}{\kappa}\Bigr)^6\Biggr) . \label{UVexp}
\end{equation}
This proves that the $\mathcal{M}_1(n,\kappa,\mu)$ approximation to $\mathcal{M}(n,\kappa,\mu)$
is valid until a few e-foldings before horizon crossing. (The graphical analysis of the previous
subsection establishes that it is actually valid for some e-foldings after horizon crossing.) 
Combining expressions (\ref{Ftrans}) and (\ref{UVexp}) also shows that the 
$\mathcal{M}_1(n,\kappa,\mu)$ approximation correctly captures the ultraviolet divergences of 
the coincident propagator and we can set $D=4$ in considering the later approximations.

The $\mathcal{M}_2(n,\kappa,\mu)$ approximation pertains after horizon crossing when the 
4th and 6th terms of (\ref{Meqn}) have red-shifted into insignificance. Dropping these terms,
and recalling the definition (\ref{freqs}) of the frequency $\omega^2(n,\mu)$ we have,
\begin{equation}
\mathcal{M}'' + \frac12 {\mathcal{M}'}^2 + (3 \!-\! \epsilon) \mathcal{M}' + \frac92 - 
2 \omega^2 \simeq 0 \; . \label{M2eqn}
\end{equation}
Only derivatives of $\mathcal{M}$ appear in equation (\ref{M2eqn}), and there is no explicit
dependence on $\kappa$. Hence $\mathcal{M}'$ can only depend on $\kappa$ through the initial
condition at $n_2 = n_{\kappa} + 4$, and Figures~\ref{kappasmall}-\ref{kappabig} reveal no
such dependence. We are therefore led to the change of variable,
\begin{equation}
\mathcal{M}'(n,\mu) \simeq -3 + 2 \omega(n,\mu) \tanh\Bigl[ \alpha(n,\mu)\Bigr] \label{alpha}
\end{equation}
Substituting (\ref{alpha}) in (\ref{M2eqn}) leads to the relation,
\begin{equation}
2 \omega \alpha' {\rm sech}^2(\alpha) - 2 \omega^2 {\rm sech}^2(\alpha) - 
\frac{9\epsilon}{2 \omega} \tanh(\alpha) + 3 \epsilon \simeq 0 \; . \label{alphaeqn}
\end{equation}
Ignoring the last 2 terms, and imposing the correct initial condition at $n = n_2$ implies,
\begin{equation}
\mathcal{M}'(n,\mu) \simeq -3 + 2 \omega \tanh\Bigl[\alpha_2 + \!\int_{n_2}^{n} \!\!\!
dn' \, \omega(n',\mu)\Bigr] \; , \; \alpha_2 = \tanh^{-1}\Bigl[ 
\frac{ \mathcal{M}_2' \!+\! 3}{2 \omega_2}\Bigr] \; . \label{alphasoln}
\end{equation}
Integrating and using the $\kappa$-dependent initial condition gives,
\begin{equation}
\mathcal{M}(n,\kappa,\mu) \simeq \mathcal{M}_2 - 3 (n \!-\! n_2) + 2 \ln\Biggl[ 
\frac{\cosh[\alpha_2 + \int_{n_2}^{n} dn' \omega(n',\mu)]}{\cosh(\alpha_2)}\Biggr] \; .
\label{M2almost}
\end{equation}
Breaking up the sum in the argument of the hyperbolic cosine leads to the 
$\mathcal{M}_2(n,\kappa,\mu)$ approximation (\ref{M2def}). Because the term that was neglected
in passing from (\ref{alphaeqn}) to (\ref{alphasoln}) diverges when $\omega(n,\mu)$ vanishes, 
one expects the approximation to break down near $n = n_{\mu}$. This is just discernible in 
Figure~\ref{mu12B}.

The $\mathcal{M}_3(n,\kappa,\mu)$ approximation pertains after $\omega^2(n,\mu) = 
-\Omega^2(n,\mu)$ has changed from positive to negative. Although the 4th term in 
(\ref{Meqn}) continues to be negligible, the 6th term becomes significant over very brief
intervals when $\mathcal{M}(n,\kappa,\mu)$ falls suddenly. Modeling this correctly is
challenging because the intervals over which the 6th term matters are so short. The simplest 
approach turns out to be ignoring 6th term, and accounting for its brief impact with a
judicious absolute value.

If $n > n_{\mu}$, and we continue to ignore the 4th and 6th terms in (\ref{Meqn}), the 
appropriate equation for $\mathcal{M}(n,\kappa,\mu)$ is,
\begin{equation}
\mathcal{M}'' + \frac12 {\mathcal{M}'}^2 + (3 \!-\! \epsilon) \mathcal{M}' + \frac92 +
2 \Omega^2 \simeq 0 \; . \label{M3eqn}
\end{equation}
The appropriate substitution is,
\begin{equation}
\mathcal{M}'(n,\mu) \simeq -3 + 2 \Omega(n,\mu) \tan\Bigl[ \beta(n,\mu)\Bigr] \label{beta}
\end{equation}
Making this substitution brings equation (\ref{M3eqn}) to the form,
\begin{equation}
2 \Omega \beta' \sec^2(\beta) + 2 \Omega^2 \sec^2(\beta) + \frac{9 \epsilon}{2 \Omega} 
\tan(\beta) + 3 \epsilon \simeq 0 \; . \label{betaeqn}
\end{equation}
If we ignore the final term then the solution is,
\begin{equation}
\mathcal{M}'(n,\mu) \simeq -3 + 2 \Omega \tan\Bigl[\beta_3 - \!\int_{n_3}^{n} \!\!\!
dn' \, \Omega(n',\mu)\Bigr] \; , \; \beta_3 = \tan^{-1}\Bigl[ \frac{ \mathcal{M}_3' \!+\! 
3}{2 \Omega_3}\Bigr] \; . \label{betasoln}
\end{equation}
Integrating this expression, and supplying the aforementioned absolute value gives,
\begin{equation}
\mathcal{M}(n,\kappa,\mu) \simeq \mathcal{M}_3 - 3 (n \!-\! n_3) + 2 \ln\Biggl[ 
\Biggl\vert \frac{\cos[\beta_3 - \int_{n_3}^{n} dn' \Omega(n',\mu)]}{\cos(\beta_3)}
\Biggr\vert \Biggr] \; . \label{M3almost}
\end{equation}
Breaking up the argument of the cosine gives expression (\ref{M3def}) for
$\mathcal{M}_{3}(n,\kappa,\mu)$.

\section{Computing the Inflaton Effective Potential}

The point of developing approximate analytic forms for $\mathcal{M}(n,\kappa,\mu)$
like expressions (\ref{M1def}), (\ref{M2def}) and (\ref{M3def}) is to compute the
derivative of the effective potential (\ref{Veffeqn}) through the coincident 
propagator (\ref{Ftrans}). That is the purpose of this section. We first decompose
the integration into ultraviolet, for which expression (\ref{M1def}) pertains, and
infrared, for which expressions (\ref{M2def}) and (\ref{M3def}) apply. Only the 
ultraviolet part requires dimensional regularization, whereas the $\kappa$ dependence
of the infrared part can be factored out, and expressed in terms of 
$\mathcal{M}_1(n,\kappa,\mu)$, using relations (\ref{kappadep2}-\ref{kappadep3}).
 
The graphical analysis of section 2.2 shows that a reasonable point for the transition
between the $\mathcal{M}_1$ and the $\mathcal{M}_2$ approximations is about 4 e-foldings 
after horizon crossing. This corresponds to a wave number $K(n)$,
\begin{equation}
K(n) \equiv \frac{e^{n-4} \chi(n\!-\!4)}{\sqrt{8\pi G}} \quad \Longrightarrow \quad
\begin{cases} 
{\rm UV:} & k > K(n) \\
{\rm IR:} & k < K(n)
\end{cases}
\; . \label{Kdef}
\end{equation}
This distinction allows us to approximate the coincident propagator (\ref{Ftrans}) as,
\begin{eqnarray}
i\Delta(x;x) & = & \frac{2 \sqrt{8\pi G}}{(4\pi)^{\frac{D-1}{2}} \Gamma(\frac{D-1}{2})}
\int_{0}^{\infty} \!\!\!\! dk \, k^{D-2} \, e^{\mathcal{M}(n,\kappa,\mu)} \; , \\
& \simeq & \frac{2 \sqrt{8\pi G}}{(4\pi)^{\frac{D-1}{2}} \Gamma(\frac{D-1}{2})}
\int_{0}^{\infty} \!\!\!\! dk \, k^{D-2} \, e^{\mathcal{M}_1(n,\kappa,\mu)} \nonumber \\
& & \hspace{2cm} + \frac{\sqrt{8\pi G}}{2 \pi^2} \int_{0}^{K(n)} \!\!\!\! dk \, k^2 
\Biggl[ e^{\mathcal{M}_{2,3}(n,\kappa,\mu)} - e^{\mathcal{M}_1(n,\kappa,\mu)}\Biggr] 
\; . \qquad \label{2ints}
\end{eqnarray}
We define the first term on the right hand side of (\ref{2ints}) as $i\Delta_1(x;x)$,
and the second term as $i\Delta_{\rm IR}(x;x)$.

We compute $i\Delta_1(x;x)$ using expression (\ref{M1def}) with integral 6.574 \#2 
of \cite{Gradshteyn:1965},
\begin{equation}
i\Delta_1(x;x) = \frac{[(1 \!-\! \epsilon) H]^{D-2}}{(4\pi)^{\frac{D}2}} \times 
\frac{\Gamma( \frac{D-1}{2} \!+\! \nu) \Gamma(\frac{D-1}{2} \!-\! \nu)}{
\Gamma(\frac12 \!+\! \nu) \Gamma(\frac12 \!-\! \nu)} \times \Gamma\Bigl(1 \!-\! \frac{D}2
\Bigr) \; , \label{Delta1A}
\end{equation}
where $\nu(n,\mu)$ is given in expression (\ref{znudef}). Writing $D = 4 - \delta$ and 
expanding key terms gives,
\begin{eqnarray}
\lefteqn{i\Delta_1(x;x) = \frac{[(1 \!-\! \epsilon) H]^{D-2}}{(4\pi)^{\frac{D}2}} 
\Bigl[ \Bigr(\frac{D \!-\! 3}{2}\Bigr)^2 - \nu^2\Bigr] } \nonumber \\
& & \hspace{3cm} \times -\frac{2}{\delta} \Biggl\{1 - \Bigl[ \psi\Bigl(\frac12 + \nu\Bigr) 
+ \psi\Bigl(\frac12 - \nu\Bigr)\Bigr] \frac{\delta}{2} + O(\delta^2) \Biggr\} , \qquad
\label{Delta1B} \\
& & \hspace{0.5cm} = \frac{[(1 \!-\! \epsilon) H]^{D-4}}{(4\pi)^{\frac{D}2}} \Biggl\{ 
-\frac14 \Bigl(\frac{D \!-\! 2}{D \!-\! 1}\Bigr) R + M^2 - \frac{\delta}{2} \Bigl(1 \!-\! 
\frac{\delta}{2} \Bigr) (1 \!-\! \epsilon)^2 H^2 \Biggr\} \nonumber \\
& & \hspace{4cm} \times \Biggl\{ -\frac{2}{\delta} + \psi\Bigl(\frac12 + \nu\Bigr) +
\psi\Bigl(\frac12 - \nu\Bigr) + O(\delta) \Biggr\} , \qquad \label{Delta1C} 
\end{eqnarray}
where $R = (D-1) (\frac{D}{2} - 2 \epsilon) H^2$ is the Ricci scalar. Comparison with
expression (\ref{Veffeqn}) reveals that the conformal and quartic counterterms are,
\begin{eqnarray}
\delta \xi & = & \frac{h^2 s^{D-4}}{(4\pi)^{\frac{D}2}} \times \frac{(D \!-\! 2)}{
4 (D \!-\! 1) (D \!-\! 4)} + \frac{h^2 \delta \xi_{\rm fin}}{32 \pi^2} \; , 
\label{deltaxi} \\
\delta \lambda & = & \frac{h^4 s^{D-4}}{(4 \pi)^{\frac{D}2}} \times 
-\frac{3}{D \!-\! 4} + \frac{h^4 \delta \lambda_{\rm fin}}{64 \pi^2} \; , 
\label{deltalambda}
\end{eqnarray}
where $s$ is the mass scale of dimensional regularization and $\delta \xi_{\rm fin}$ 
and $\delta \lambda_{\rm fin}$ represent arbitrary finite renormalizations. 
Substituting expressions (\ref{Delta1C}-\ref{deltalambda}) into (\ref{Veffeqn}) and
taking the unregulated limit gives the $\mathcal{M}_1(n,\kappa,\mu)$ contribution to
the effective potential,
\begin{eqnarray}
\lefteqn{ \Bigl( \frac{\partial V_{\rm eff}}{\partial \varphi}\Bigr)_{1} = 
\frac{h^2 \varphi}{32 \pi^2} \Biggl\{ (1 \!-\! \epsilon)^2 H^2 + \delta \xi_{\rm fin} 
R + \frac1{12} \delta \lambda_{\rm fin} h^2 \varphi^2  } \nonumber \\
& & \hspace{.5cm} + \Bigl[-\frac16 R + \frac12 h^2 \varphi^2\Bigr] \Biggl[ \psi\Bigl(
\frac12 \!+\! \nu\Bigr) + \psi\Bigl(\frac12 \!-\! \nu\Bigr) + \ln\Bigl[\frac{(1 \!-\! 
\epsilon)^2 H^2}{s^2} \Bigr]\Biggr] \Biggr\} , \qquad \label{Veffprime1}
\end{eqnarray}
where the unregulated limit of the index is,
\begin{equation}
\nu = \frac{\sqrt{(\frac{3 - \epsilon}{2})^2 - z }}{1 \!-\! \epsilon} \qquad , \qquad 
z \equiv \frac{h^2 \varphi^2}{2 H^2} \; . 
\label{nuunreg}
\end{equation}
Integrating with respect to $\varphi$ allows us to express the $\mathcal{M}_1$
contribution in terms of the instantaneous values of $H$, $\epsilon$ and 
$z$,
\begin{eqnarray}
\lefteqn{\Bigl( V_{\rm eff}\Bigr)_1 = } \nonumber \\
& & \hspace{-0.5cm} \frac{H^4}{32 \pi^2} \Biggl\{ (1 \!-\! \epsilon)^2 z + 
6 \delta \xi_{\rm fin} (2 \!-\! \epsilon) z + \frac{\delta \lambda_{\rm fin} z^2}{12}
-\Bigl[(2 \!-\! \epsilon) z - \frac{z^2}{2}\Bigr] \ln\Bigl[\frac{(1 \!-\! \epsilon)^2 
H^2}{s^2} \Bigr] \nonumber \\
& & \hspace{-0.5cm} + \!\! \int_{0}^{z} \!\!\! dx \, (-2 \!+\! \epsilon \!+\! x) 
\Biggl[ \psi\Biggl( \!\frac12 + \frac{\sqrt{(\frac{3 - \epsilon}{2})^2 - x}}{1 \!-\! 
\epsilon} \Biggr) \!+ \psi\Biggl( \!\frac12 - \frac{\sqrt{( \frac{3 - \epsilon}{2})^2 - x}}{1 
\!-\! \epsilon} \Biggr) \Biggr] \! \Biggr\} . \qquad \label{Veff1}
\end{eqnarray}

It remains to evaluate expression (\ref{2ints}) for $i\Delta_{\rm IR}(x;x)$. Note 
from relations (\ref{kappadep2}-\ref{kappadep3}), and the slow roll approximation for 
the post-horizon amplitude, that dependence on the mass parameter $\mu$ factors out of 
the integration over wave number $k$,
\begin{eqnarray}
i\Delta_{\rm IR}(x;x) & = & \frac{\sqrt{8\pi G}}{2 \pi^2} \int_{0}^{K(n)} \!\!\!\! dk \, 
k^2 e^{\mathcal{M}_1(n_{\kappa}+4,\kappa,\mu)} \times \Bigl\{ e^{f_{2,3}(n,\mu)} - 
e^{f_1(n,\kappa,\mu)} \Bigr\} \; , \qquad \label{DeltaIRA} \\
& \simeq & \frac1{4\pi^2} \int_{0}^{K(n)} \!\!  \frac{dk}{k} H^2(t_k) \times 
\Bigl\{ e^{f_{2,3}(n,\mu)} - e^{f_1(n,\kappa,\mu)} \Bigr\} \; , \label{DeltaIRB} 
\end{eqnarray}
where $f_i \equiv \mathcal{M}_i(n,\kappa,\mu) - \mathcal{M}_1(n_{\kappa}\!+\!4,\kappa,\mu)$.
Now change variables from $k$ to $n_{\kappa}$ using $dk/k = (1 - \epsilon) dn_{\kappa}$,
\begin{equation}
i\Delta_{\rm IR}(x;x) \simeq \frac1{4\pi^2} \int_{0}^{n-4} \!\!\!\! dn_{\kappa} 
\frac{[1 \!-\! \epsilon(n_{\kappa})] \chi^2(n_{\kappa})}{8 \pi G} \times 
\Bigl\{ e^{f_{2,3}(n,\mu)} - e^{f_1(n,e^{n_{\kappa}} \chi(n_{\kappa}),\mu)} \Bigr\} 
\; . \label{DeltaIRC}
\end{equation}
Note the distinction between the integration over $n_{\kappa}$ and the multiplicative 
factor which depends on the local e-folding $n$. This means that $i\Delta_{\rm IR}(x;x)$
is not even a local function of the inflationary geometry. The factors of $e^{f_2(n,\mu)}$
and $e^{f_3(n,\mu)}$ also depend nonlocally on the geometry,
\begin{eqnarray}
e^{f_2(n,\mu)} & = & e^{-3 (n - n_2)} \frac{ \cosh^2[\alpha_2 + \int_{n_2}^{n} dn' 
\omega(n',\mu)]}{\cosh^2(\alpha_2)} \; , \label{expf2} \\
e^{f_3(n,\mu)} & = & e^{f_2(n,\mu)} \times \frac{ \cos^2[\beta_3 - \int_{n_3}^{n} dn' 
\Omega(n',\mu)]}{\cos^2(\beta_3)} \; . \label{expf3}
\end{eqnarray}

Expression (\ref{DeltaIRC}) gives the nonlocal contribution to the coincident propagator.
To find the corresponding contribution to the effective potential one substitutes this 
in expression (\ref{Veffeqn}) and then integrates with respect to the inflaton field 
$\varphi$, not excepting the dependence on $\mu^2 = 8\pi G \times \frac12 h^2 \varphi^2$. 
Note that although the result depends nonlocally on the geometry, it is local in the 
inflaton. Obtaining an explicit result is not possible owing to the complicated 
$\mu$-dependence of the functions $f_2(n,\mu)$ and $f_3(n,\mu)$, however, under the 
(expected) assumption that expressions (\ref{expf2}-\ref{expf3}) are small, the result 
is,
\begin{equation}
\Bigl( V_{\rm eff}\Bigr)_{IR} \simeq \frac14 h^2 \varphi^2 \times -\frac1{4\pi^2} 
\int_{0}^{n-4} \!\!\!\! dn' [1 \!-\! \epsilon(n')] H^2(n') \; . \label{VeffIR}
\end{equation}
We recognize (\ref{VeffIR}) as a negative contribution to the inflaton mass-squared.
How significant it is depends on the classical model of inflation. For the quadratic 
potential we have been assuming, with our initial conditions, the correction 
(\ref{VeffIR}) would subtract off a fraction $\frac{75 h^2}{16 \pi^2} \simeq \frac12 
h^2$ of the inflaton mass. This could make significant changes to the inflaton's 
evolution, but these could be compensated by increasing the bare mass. The 
consequences for the geometry are more difficult to estimate but two features are 
obvious:
\begin{itemize}
\item{Nonlocal corrections can contaminate late time physics, when scales should 
be small, with information from very early times, when scales were large; and}
\item{Nonlocal corrections cannot be subtracted off using local actions.}
\end{itemize}

\section{Modified Friedmann Equations}

The purpose of this section is to derive the nontrivial equations for a gravitational 
Lagrangian whose specialization to the geometry of inflation (\ref{FLRW}) takes the 
form $L(a,\dot{a},\ddot{a}) = a^3 f(H,\epsilon)$. If we knew the Lagrangian for a 
general metric then first varying with respect to $g^{\mu\nu}$ and afterwards 
specializing to (\ref{FLRW}) would give two nontrivial equations, one from the
variation with respect to $g^{ij}$ and the other from the variation with respect to
$g^{00}$. However, specializing the geometry first gives only one equation. The famous
theorem of Palais \cite{Palais:1979rca} assures us that this equation is correct, and
we know from the fact that $g_{ij} = a^2 \delta_{ij}$ that this equation is proportional 
to the variation with respect to $g^{ij}$. We begin by deriving this equation, then we
reconstruct the $g^{00}$ equation using conservation. The section closes by checking 
that our results agree for the special cases of no dependence upon $\epsilon$ and also
$F(R)$ models.

For a Lagrangian of the form $L(a,\dot{a},\ddot{a}) = a^3 f(H,\epsilon)$ the Euler-Lagrange
equation is,
\begin{eqnarray}
\lefteqn{ \frac{\partial L}{\partial a} - \frac{d}{dt} \Bigl( \frac{\partial L}{
\partial \dot{a}}\Bigr) + \frac{d^2}{dt^2} \Bigl( \frac{\partial L}{\partial \ddot{a}}
\Bigr) = a^2 \Bigl[ 3f - H \frac{\partial f}{\partial H} - (1 \!-\! \epsilon) 
\frac{\partial f}{\partial \epsilon}\Bigr] } \nonumber \\
& & \hspace{2.5cm} - \frac{d}{dt} \Biggl\{ a^2 \Bigl[ \frac{\partial f}{\partial H} +
\frac{2 (1 \!-\! \epsilon)}{H} \frac{\partial f}{\partial \epsilon}\Bigr] \Biggr\} +
\frac{d^2}{dt^2} \Biggl\{ a^2 \Bigl[-\frac{1}{H^2} \frac{\partial f}{\partial \epsilon}
\Bigr] \Biggr\} , \quad \\
& & \hspace{-.2cm} = a^2 \Biggl[ 3f \!-\! \Bigl( \frac{d}{dt} \!+\! 3 H\Bigr) 
\frac{\partial f}{\partial H} \!-\! 3 (3 \!+\! \epsilon) \frac{\partial f}{\partial 
\epsilon} \!-\! 2 (3 \!+\! \epsilon) \frac1{H} \frac{d}{d t} \frac{\partial f}{\partial 
\epsilon} \!-\! \frac1{H^2} \frac{d^2}{dt^2} \frac{\partial f}{\partial \epsilon} \Biggr] . 
\quad \label{F2zero}
\end{eqnarray}
Because $g_{ij} = a^2 \delta_{ij}$ for the geometry (\ref{FLRW}) we recognize the
$g^{ij}$ equation as, 
\begin{equation}
f \!-\! \Bigl( \frac13 \frac{d}{dt} \!+\! H\Bigr) 
\frac{\partial f}{\partial H} \!-\! (3 \!+\! \epsilon) \frac{\partial f}{\partial 
\epsilon} \!-\! \Bigl(2 \!+\! \frac{2}{3} \epsilon\Bigr) \frac1{H} \frac{d}{d t} 
\frac{\partial f}{\partial \epsilon} \!-\! \frac1{3 H^2} \frac{d^2}{dt^2} 
\frac{\partial f}{\partial \epsilon} = 0 \; . \label{Feqn2}
\end{equation}
The conservation of stress-energy implies that (\ref{Feqn2}) and the missing $g^{00}$ 
equation obey,
\begin{equation}
3 H \Biggl\{ \Bigl( g^{ij} \; {\rm Eqn}\Bigr) + \Bigl( g^{00} \; {\rm Eqn}\Bigr) \Biggr\} 
= \frac{d}{d t} \Bigl( g^{00} \; {\rm Eqn}\Bigr) \; . \label{conservation}
\end{equation}
We also want the $g^{00}$ equation to contain one fewer time derivative than (\ref{Feqn2}).
A little thought reveals the solution to be,
\begin{equation}
-f + H \frac{\partial f}{\partial H} + \Bigl[3 \!+\! \epsilon + \frac1{H} 
\frac{d}{d t} \Bigr] \frac{\partial f}{\partial \epsilon} = 0 \; . \label{Feqn1}
\end{equation}
Relations (\ref{Feqn1}) and (\ref{Feqn2}) are the desired generalizations of the first and
second Friedmann equations, respectively.

Relations (\ref{Feqn1}) and (\ref{Feqn2}) obey two important correspondence limits. The
first comes from assuming that there is no dependence on $\epsilon$, as was considered in
a previous study \cite{Liao:2018sci}. When $\frac{\partial f}{\partial \epsilon} = 0$ our 
Friedmann equations agree with relations (15) and (14) from that study. The second limit 
is relevant to $F(R)$ models,
\begin{equation}
f(H,\epsilon) \longrightarrow F\Bigl( (12 \!-\! 6 \epsilon) H^2 \Bigr) \; . 
\label{FR}
\end{equation}
In that limit we have,
\begin{equation}
\frac{\partial f}{\partial H} \longrightarrow (24 \!-\! 12 \epsilon) H \!\times\! F'
\qquad , \qquad \frac{\partial f}{\partial \epsilon} \longrightarrow -6 H^2 \!\times\! F'
\; . \label{FRlimit}
\end{equation}
Substituting these relations into our first Friedmann equation (\ref{Feqn1}) gives,
\begin{equation}
\Bigl( {\rm Eqn\ \ref{Feqn1}}\Bigr) \longrightarrow -F(R) + 6 (1 \!-\! \epsilon) H^2 F'(R)
- 6 H \frac{d}{d t} F'(R) = 0 \; . \label{FR1}
\end{equation}
With (\ref{FRlimit}) our second Friedmann equation (\ref{Feqn2}) becomes,
\begin{equation}
\Bigl( {\rm Eqn\ \ref{Feqn2}}\Bigr) \longrightarrow + F(R) - (6 \!-\! 2 \epsilon) H^2 F'(R)
+ 4 H \frac{d}{dt} F'(R) + 2 \frac{d^2}{d t^2} F'(R) = 0 \; . \label{FR2}
\end{equation}
Relations (\ref{FR1}-\ref{FR2}) can be recognized as the specialization the $F(R)$ field
equations to the geometry (\ref{FLRW}) of inflation.

The Friedmann equation of general relativity ($3 H^2 = 8\pi G \rho$) involves only 
first derivatives of the scale factor $a(t)$, whereas our generalization (\ref{Feqn1}) 
involves {\it three} derivatives. Similarly, the $g_{ij}$ equation of general 
relativity ($-2 \dot{H} -3 H^2 = 8\pi G p$) involves second derivatives of $a(t)$, 
whereas our generalization (\ref{Feqn2}) involves {\it four} time derivatives. 
Classical theories that involve higher time derivatives have new degrees of freedom
which are typically kinetically unstable \cite{Woodard:2006nt}. The unique local and
invariant modification of general relativity that avoids kinetic instabilities is
$F(R)$ models (\ref{FR}) \cite{Woodard:2006nt}, but a glance at expression 
(\ref{Veff1}) reveals that the myriad factors of $H$ and $\epsilon$ in the effective 
potential are not limited to the combination $R = 6 (2 - \epsilon) H^2$. This sounds
like a major problem, and it probably is, but not in the direct way one might think.
Quantum corrections to the effective action are typically not even local, and yet 
they introduce no new degrees of freedom nor any essential instability. The right way 
to understand these higher derivative or nonlocal quantum corrections is as {\it 
perturbations} to the existing solutions of the classical, lower-derivative theory 
\cite{Simon:1990ic}. Quantum corrections introduce no new degrees of freedom, they 
simply distort the evolution of the classical degrees of freedom. Unfortunately, the 
distortion from cosmological Coleman-Weinberg potentials is typically too large 
because these corrections are not Planck-suppressed. To avoid large distortions one
must subtract most of the cosmological Coleman-Weinberg potential using a classical 
modification of the original model. Because the subtraction has the status of a
modification to the classical action, it can induce new degrees of freedom and 
instabilities. A recent study of subtractions involving functions of the inflaton and
the Ricci scalar reveals that the higher derivative degrees of freedom cause 
inflation to end after an infinitesimal number of e-foldings \cite{Miao:2019bnq}.

\section{Epilogue}

We have developed an analytic approximation for the logarithm of the amplitude of the
norm-squared mode function for a massive, minimally coupled scalar in the presence of
an arbitrary inflating background. Our result takes the form of a sequence of approximate
forms that apply when the physical wave number dominates the Hubble parameter --- 
expression (\ref{M1def}) --- when the Hubble parameter dominates the physical wave 
number and the mass --- expression (\ref{M2def}) --- and after the mass dominates the 
Hubble parameter --- expression (\ref{M3def}). Section 2.2 contains many graphs which 
demonstrate the validity of these approximations for inflation with a quadratic 
potential, and section 2.3 gives analytic derivations which should apply for general
models.

\begin{figure}[H]
\includegraphics[width=4.5cm,height=4cm]{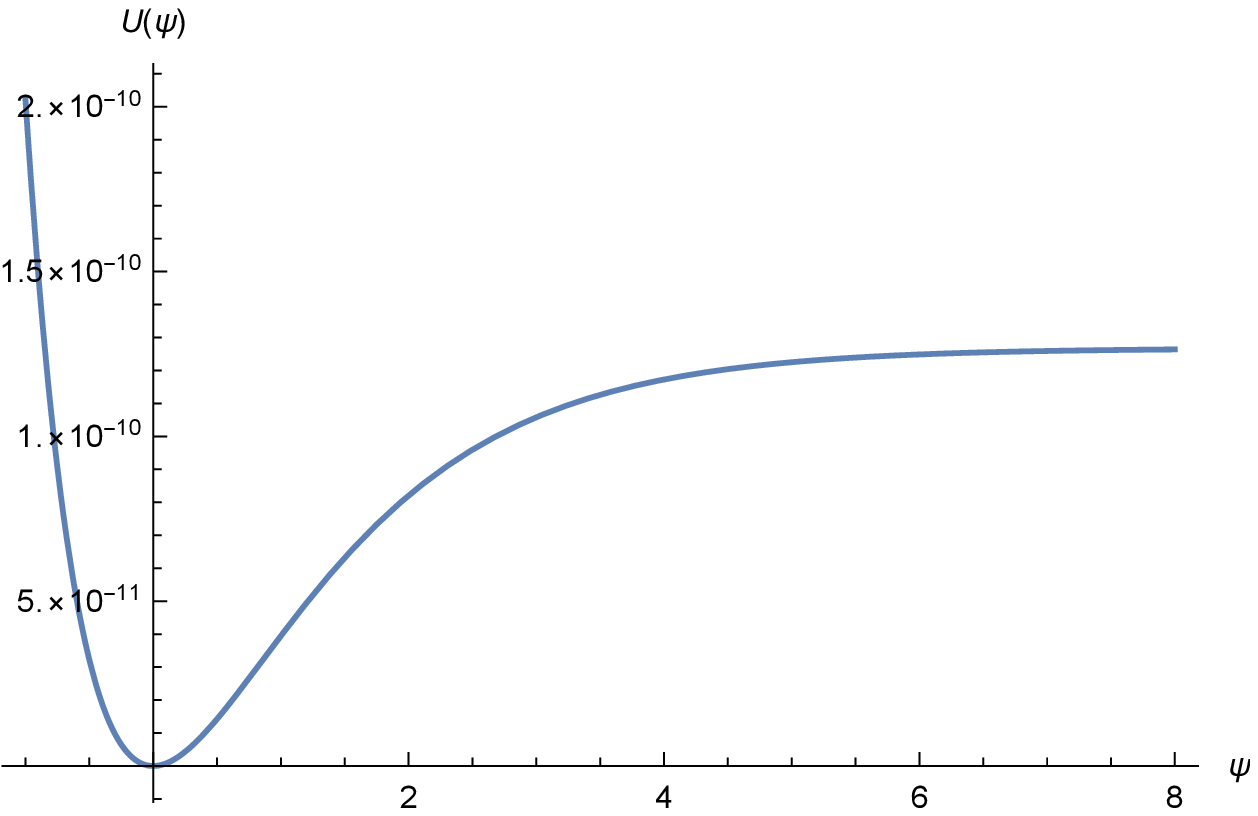}
\hspace{-0.1cm}
\includegraphics[width=4.5cm,height=4cm]{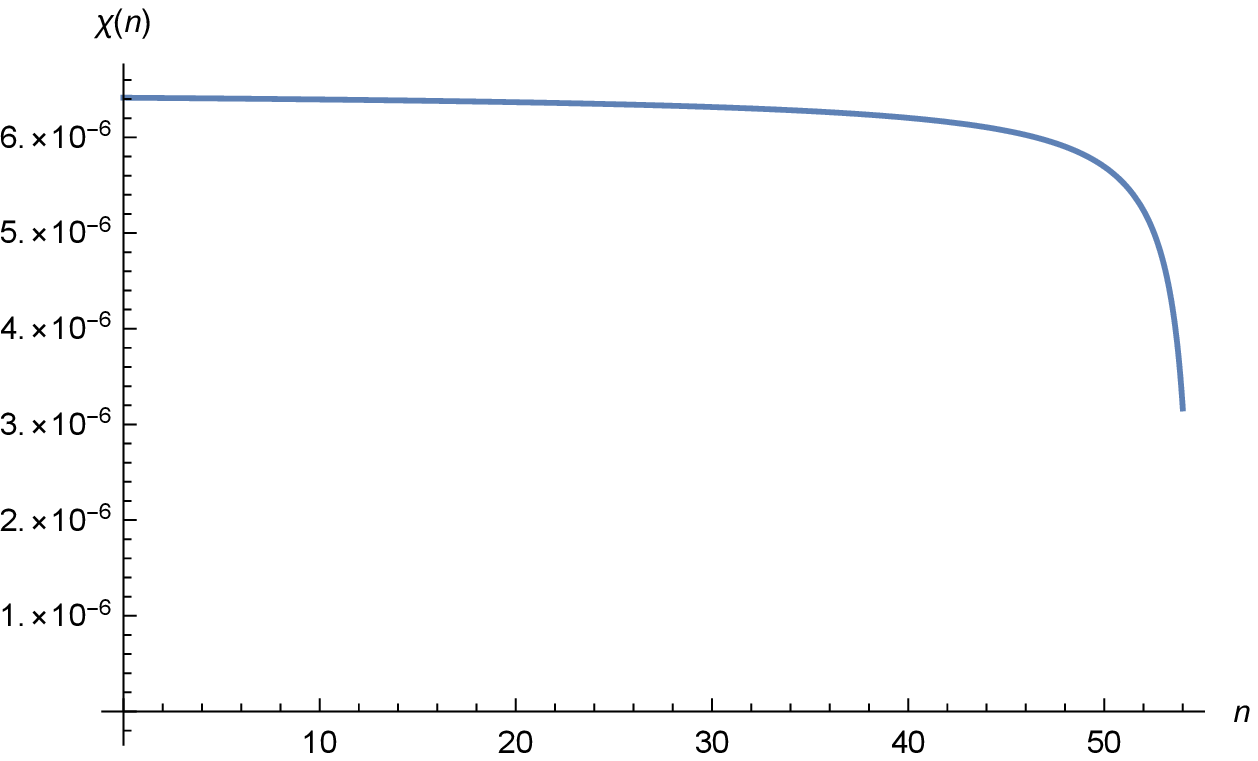}
\hspace{-0.1cm}
\includegraphics[width=4.5cm,height=4cm]{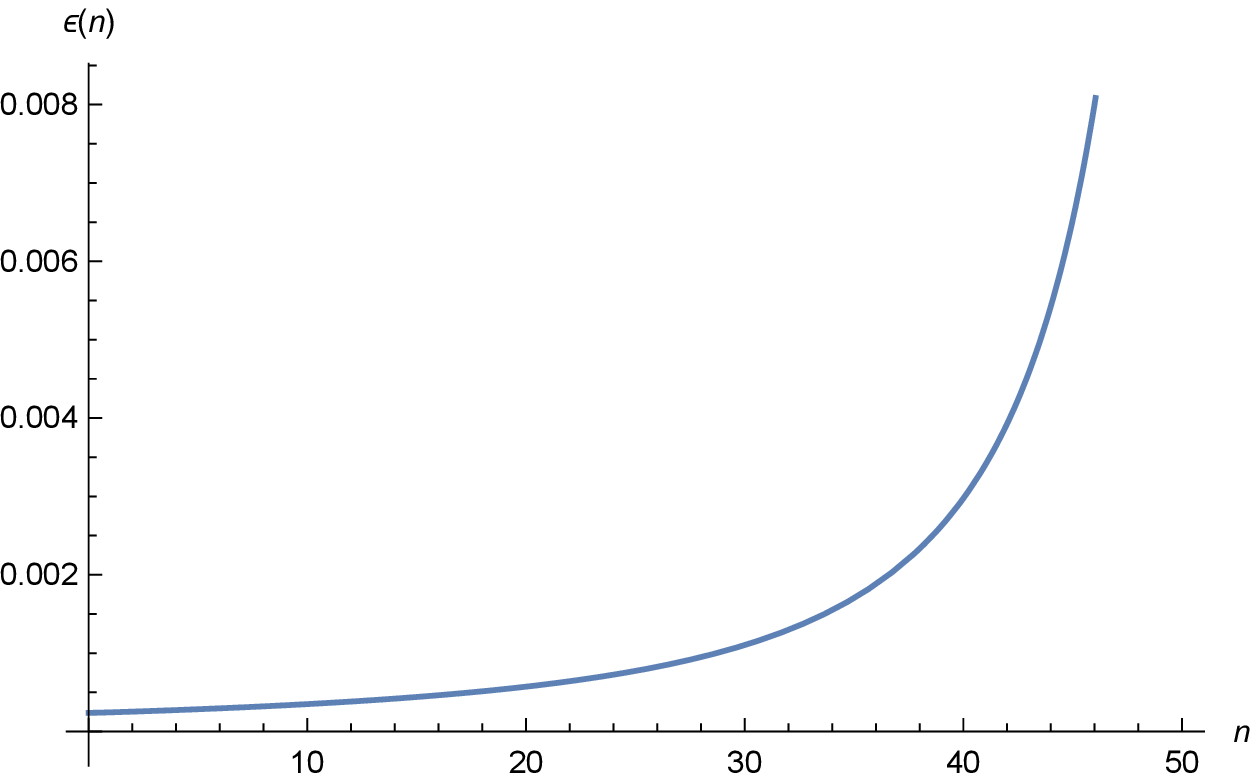}
\caption{\footnotesize{These graphs show the Starobinsky potential $U(\psi)$ of 
expression (\ref{StaroU}), as well as the dimensionless Hubble parameter $\chi(n)$ 
and the first slow roll parameter $\epsilon(n)$ for inflation starting from 
$\psi_0 = 5.3$.}}
\label{Starobinsky1}
\end{figure}

The quadratic dimensionless potential $U(\psi) = \frac12 c^2 \psi^2$ was chosen for 
our detailed studies because the slow roll approximations (\ref{slowroll}) give simple, 
analytic expressions for the dimensionless Hubble parameter $\chi(n)$ and the first 
slow roll parameter $\epsilon(n)$. With the choice of $c \simeq 7.1 \times 10^{-6}$ this 
model is consistent with observations of the scalar amplitude and the scalar spectral 
index \cite{Aghanim:2018eyx}. However, the model is excluded by its high prediction of 
$r \simeq 0.14$ for the tensor-to-scalar ratio \cite{Aghanim:2018eyx}. It is worth
briefly considering how our analysis applies to the plateau potentials that are 
currently permitted by the data. Perhaps the simplest of these is the Einstein-frame
version of the model proposed by Starobinsky \cite{Starobinsky:1980te}, whose
dimensionless potential is \cite{Brooker:2016oqa},
\begin{equation}
U(\psi) = \frac34 M^2 \Bigl( 1 - e^{-\sqrt{\frac23} \, \psi}\Bigr)^2 \qquad , \qquad
M^2 = 1.3 \times 10^{-5} \; . \label{StaroU}
\end{equation}
Starting from $\psi_0 = 5.3$ gives a little over 50 e-foldings of inflation, and
the model is consistent with observations. A glance at Figure~\ref{Starobinsky1}
reveals how this is achieved: the dimensionless Hubble parameter $\chi(n)$ is almost 
constant, implying a very small value of the first slow roll parameter $\epsilon(n)$.  

\begin{figure}[H]
\includegraphics[width=4.5cm,height=4cm]{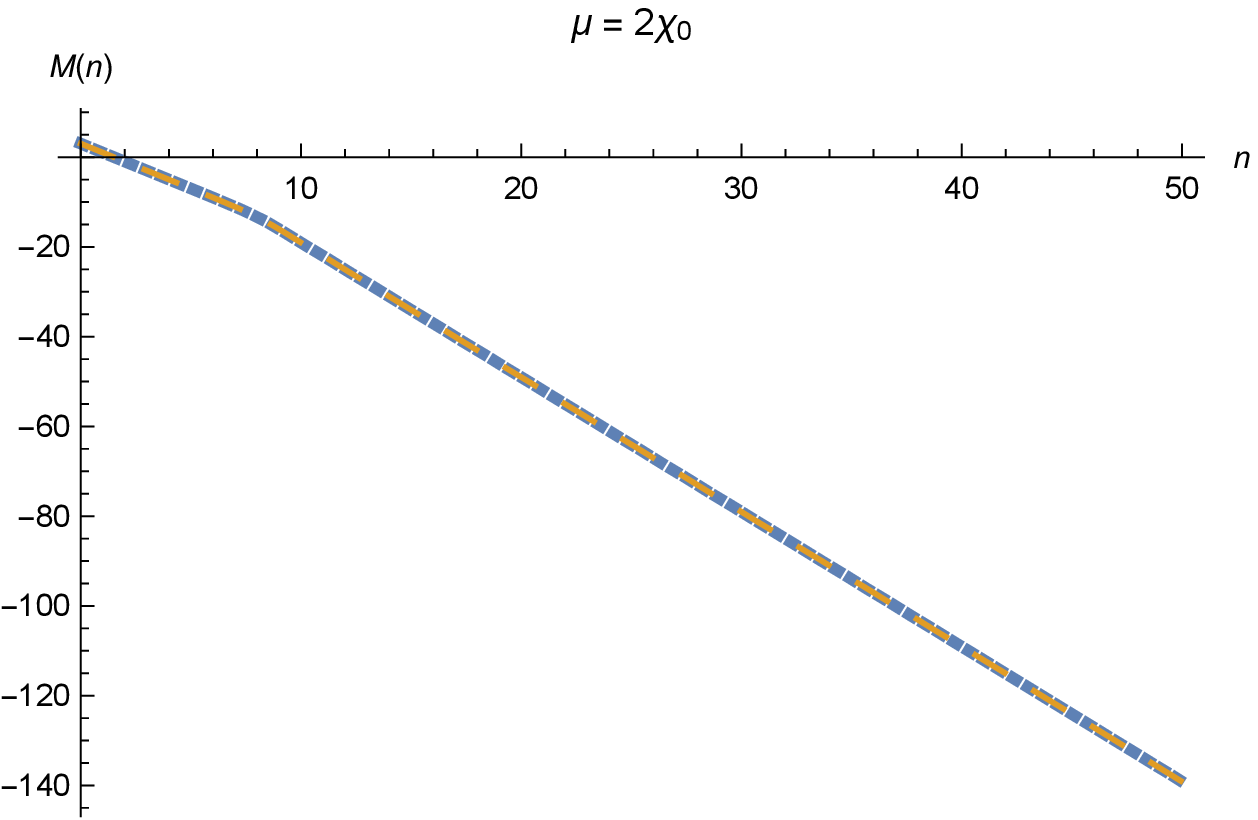}
\hspace{-0.1cm}
\includegraphics[width=4.5cm,height=4cm]{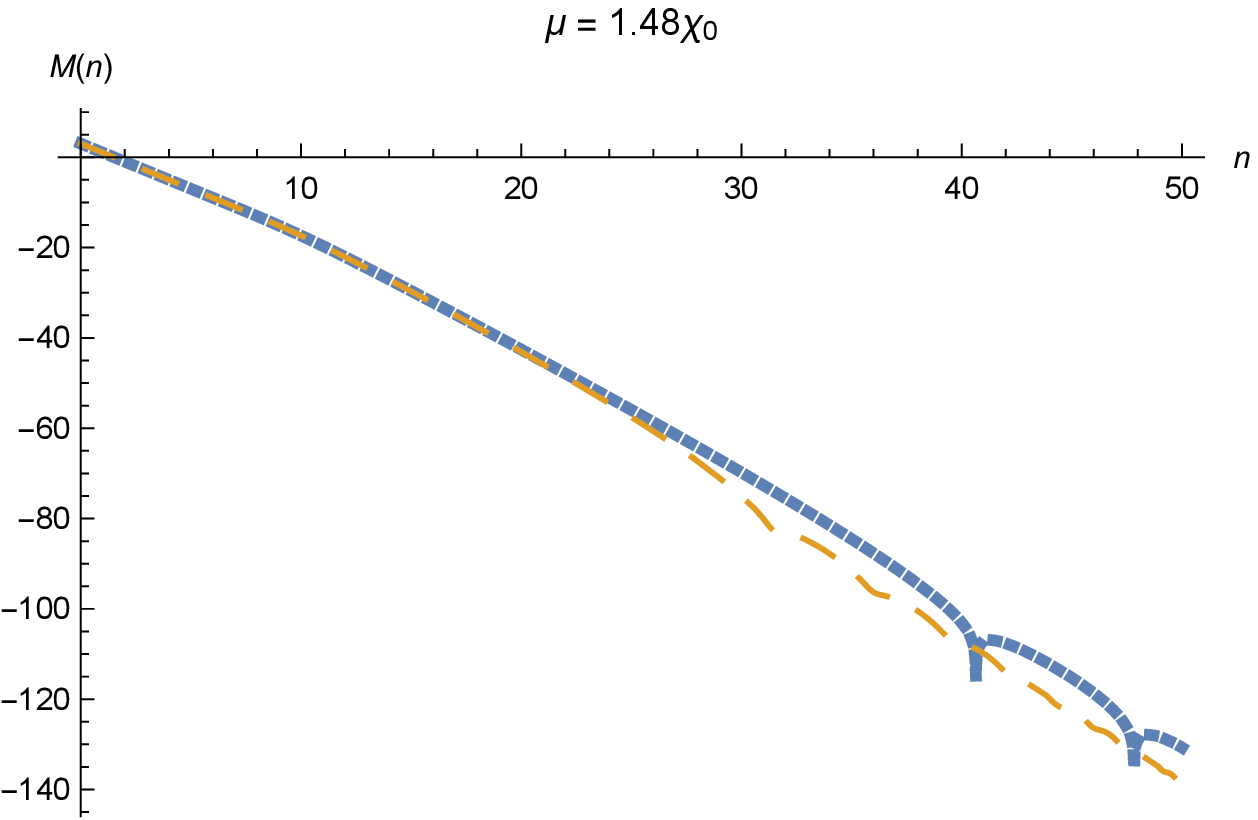}
\hspace{-0.1cm}
\includegraphics[width=4.5cm,height=4cm]{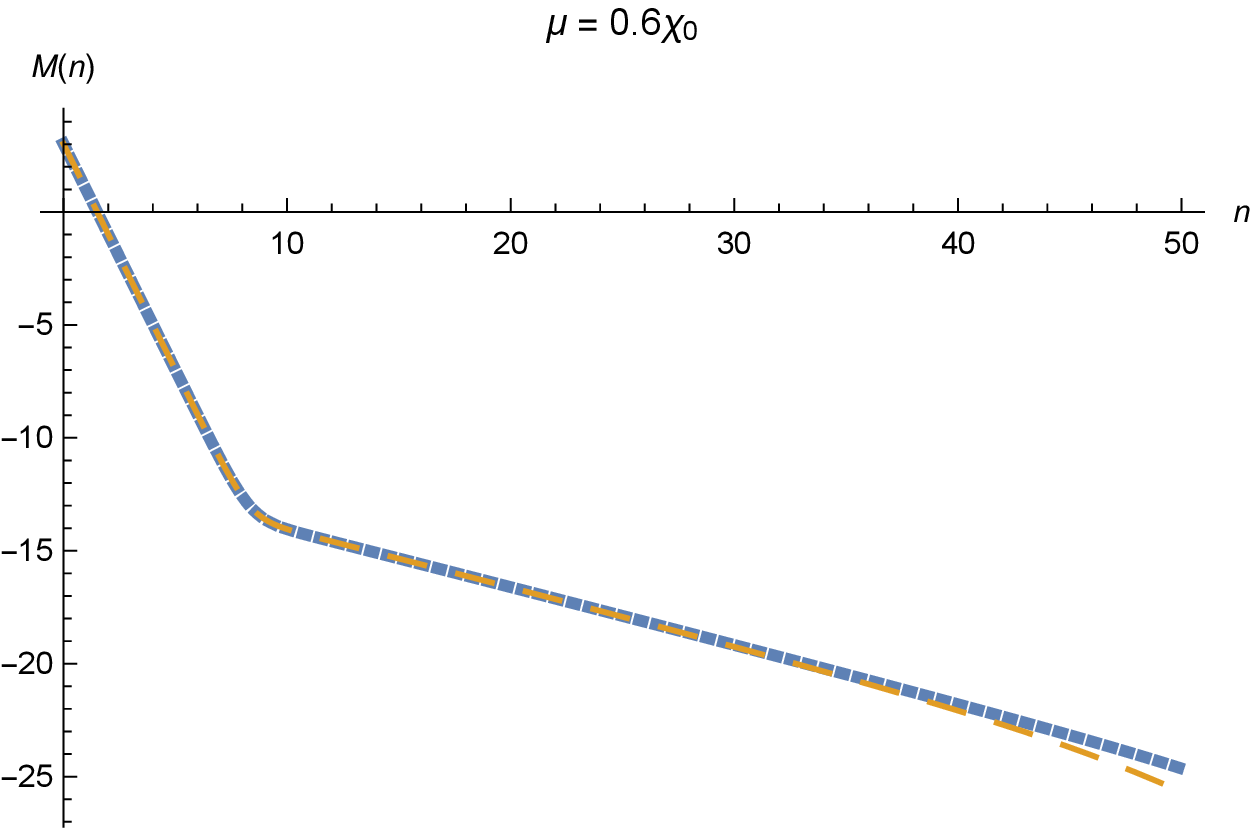}
\caption{\footnotesize{These graphs compare the numerical solution of 
$\mathcal{M}(n,\kappa,\mu)$ (in blue dots) with the $\mathcal{M}_1(n,\kappa,\mu)$
approximation (in long yellow dashes) for inflation driven by the Starobinsky
potential (\ref{StaroU}).}}
\label{Starobinsky2}
\end{figure}

All our approximations continue to apply to this model, but the extreme
flatness of $\chi(n)$ restricts the range of dimensionless mass parameters $\mu$
for which $\chi(n)$ can make the transition from being greater than $\frac23 \mu$ 
to being less that is associated with the $\mathcal{M}_3(n,\kappa,\mu)$
approximation (\ref{M3def}). This is shown in Figure~\ref{Starobinsky2}, which
compares $\mathcal{M}(n,\kappa,\mu)$ with the $\mathcal{M}_1(n,\kappa,\mu)$
approximation (\ref{M1def}) for three different values of $\mu$.  
\noindent When $\mu = 2 \chi_0$ the dimensionless Hubble parameter is always less
than $\frac23 \mu$ and the $\mathcal{M}_1(n,\kappa,\mu)$ approximation remains valid
throughout inflation. Indeed, the graph for $\mu = 2 \chi_0$ is almost identical to 
the $\mu = 2 \chi_0$ graph for the quadratic model in Figure~\ref{mu10and2}. Only 
when $\mu$ is slightly smaller than $\frac32 \chi_0$ can the transition be made to 
happen during inflation, as it does for the case of $\mu = 1.48 \chi_0$ in 
Figure~\ref{Starobinsky2}. For this case the transition comes at $n_{\mu} \simeq 
28.2$, and one can see from the graph that all three approximations are required.
When $\mu$ becomes even slightly smaller the transition does not occur during 
inflation. For example, the case of $\mu = 1.4 \chi_0$ corresponds to $n_{\mu} 
\simeq 46.7$. For smaller values of $\mu$ the $\mathcal{M}_1(n,\kappa,\mu)$ 
approximation applies until somewhat after horizon crossing, after which the
$\mathcal{M}_2(n,\kappa,\mu)$ approximation pertains. The case of $\mu = 0.6 
\chi_0$ is given in Figure~\ref{Starobinsky2}, and is similar to what 
Figure~\ref{mu06} shows for the quadratic model with the same value of $\mu$. 

The primary motivation for this work was to determine how Coleman-Weinberg corrections
to the inflaton potential depend on the geometry of inflation so that previous studies 
of their effects \cite{Liao:2018sci,Miao:2019bnq} can be extended. Our result consists of
a local part (\ref{Veff1}) that depends on the instantaneous values of $H$ and $\epsilon$,
and a nonlocal part --- derived from integrating $\frac12 h^2 \varphi$ times expression 
(\ref{DeltaIRC}) --- that depends on the past geometry. {\it We emphasize that these
approximations are independent of the classical potential and depend only on the coupling
(\ref{scalarcoupling}) between the inflaton $\varphi$ and the scalar $\phi$.} It is also
significant that the various factors of $H$ and $\epsilon$ in the local part (\ref{Veff1}) 
are not restricted to the Ricci scalar $R = 6 (2 - \epsilon) H^2$. This fact, and the 
existence of the nonlocal correction, were predicted on the basis of indirect arguments 
\cite{Miao:2015oba}. 

Because the most general, stable subtraction is a local function of $\varphi$ and $R$,
{\it complete subtraction is impossible and there can no longer be any doubt that 
cosmological Coleman-Weinberg potentials make significant changes to inflation.} 
Determining what those changes are requires understanding how cosmological Coelman-Weinberg
potentials modify the Friedmann equations. Generalizing the Friedmann equations to include 
the nonlocal contributions is challenging, but those contributions should be small because
the mass term suppresses the amplitude at late times. The Friedmann equations
appropriate to the local part (\ref{Veff1}) are (\ref{Feqn1}) and (\ref{Feqn2}).

Although we have considered cosmological Coleman-Weinberg potentials from coupling
the inflaton to a scalar (\ref{scalarcoupling}), our results should be easily extendable
to more general couplings. So we will finally be able to extend the old de Sitter 
results to a general inflationary background for the case of an inflaton which is 
Yukawa-coupled to fermions \cite{Candelas:1975du,Miao:2006pn}, and to a charged 
inflaton which is coupled to a vector boson \cite{Allen:1983dg,
Prokopec:2007ak}.\footnote{This has just been accomplished for the fermionic case 
\cite{Sivasankaran:2020dzp}.} It is also worth pointing out that our approximation 
seems to be valid even for a moderately time dependent inflaton $\varphi_0(t)$. Indeed, 
the crucial $\mathcal{M}_1$ approximation (\ref{M1def}) was originally motivated by the 
exact result for $\phi_0(t) \propto H(t)$ \cite{Janssen:2009pb} for constant $\epsilon$.

Finally, we should mention the impact of cosmological Coleman-Weinberg potentials on 
primordial perturbations, which are the principal observable from inflation. Slow roll
approximations for the scalar and tensor power spectra can be expressed as functions 
of the e-folding $n$ at which a perturbation experiences horizon crossing, in terms of 
the dimensionless Hubble parameter $\chi(n)$ and first slow roll parameter $\epsilon(n)$,
\begin{equation}
\Delta^2_{\mathcal{R}}(n) \simeq \frac1{8 \pi^2} \!\times\! \frac{\chi^2(n)}{\epsilon(n)}
\qquad , \qquad \Delta^2_{h}(n) \simeq \frac1{8 \pi^2} \!\times\! 16 \chi^2(n) \; .
\label{spectra}
\end{equation}
The analogous slow roll approximations for the scalar spectral index and the 
tensor-to-scalar ratio are,
\begin{equation}
1 - n_s(n) \simeq 2 \epsilon(n) + \frac{\epsilon'(n)}{\epsilon(n)} \qquad , \qquad
r(n) \simeq 16 \epsilon(n) \; . \label{indices}
\end{equation}
Arbitrarily accurate analytic approximations are available \cite{Brooker:2017kjd,
Brooker:2017kij} if greater precision is required, and the same is true for the
tri-spectrum of non-Gaussianity \cite{Basu:2019jew}.

Cosmological Coleman-Weinberg potentials change the predictions 
(\ref{spectra}-\ref{indices}) by changing the numerical values of the geometrical 
parameters $\chi(n)$ and $\epsilon(n)$ through the modified Friedmann equations 
(\ref{Feqn1}) and (\ref{Feqn2}). Unless the coupling constant $h$ is made so small 
as to preclude efficient re-heating, the changes induced by our result (\ref{Veff1}) 
are far too large for any viable classical model of inflation. This has long been 
apparent from the flat space limit $(V_{\rm eff})_1 \rightarrow \frac{h^4 \varphi^4}{
256 \pi^2} \ln(\frac{h^2 \varphi^2}{2 s^2})$ \cite{Coleman:1973jx}, which is also the
large field limit. That term could be subtracted off, but the remainder after even 
the best possible subtractions still causes inflation to end too quickly 
\cite{Liao:2018sci,Miao:2019bnq}. A more promising approach seems to be arranging 
cancellations between the positive cosmological Coleman-Weinberg potentials induced 
by coupling to bosonic fields and the negative potentials induced by coupling to 
fermions \cite{Miao:2020zeh}, although no solution has been devised yet.

If an acceptable cancellation can be found it will also be necessary to check for
changes to the functional forms (\ref{spectra}-\ref{indices}) of the inflationary
observables. Because these results derive from the linearized field equations of
perturbations about the cosmological background, they might show significant changes 
even if the Bose-Fermi cancellation kept changes to the background small. One would 
need to compute the 1PI (one-particle-irreducible) 2-point functions for the 
inflaton and for the metric in an inflationary background. Those computations would 
be challenging but it is encouraging that the results for de Sitter background have 
been obtained \cite{Kahya:2007bc,Kahya:2007cm,Tsamis:1996qk}.

\vskip 1cm

\centerline{\bf Acknowledgements}

We are grateful for discussions and correspondence with A. Starobinsky.
This work was partially supported by Taiwan MOST grant 107-2119-M-006-014
and 108-2112-M-006-004; by the European Union's Seventh Framework Programme 
(FP7-REGPOT-2012-2013-1) under grant agreement number 316165; by the European 
Union's Horizon 2020 Programme under grant agreement 669288-SM-GRAV-ERC-2014-ADG;
by NSF grants PHY-1806218 and PHY-1912484; and by the Institute 
for Fundamental Theory at the University of Florida.

\end{document}